\documentclass[sigconf]{acmart}

\copyrightyear{2021}
\acmYear{2021}
\setcopyright{acmlicensed}\acmConference[SIGIR '21]{Proceedings of the 44th International ACM SIGIR Conference on Research and Development in Information Retrieval}{July 11--15, 2021}{Virtual Event, Canada}
\acmBooktitle{Proceedings of the 44th International ACM SIGIR Conference on Research and Development in Information Retrieval (SIGIR '21), July 11--15, 2021, Virtual Event, Canada}
\acmPrice{15.00}
\acmDOI{10.1145/3404835.3462897}
\acmISBN{978-1-4503-8037-9/21/07}
\usepackage{acronym}
\usepackage{booktabs}
\usepackage[skip=0pt]{caption}
\usepackage{subcaption}
\usepackage{multirow}
\usepackage[inline]{enumitem}
\usepackage{pifont}

\AtBeginDocument{%
  \providecommand\BibTeX{{%
    \normalfont B\kern-0.5em{\scshape i\kern-0.25em b}\kern-0.8em\TeX}}}

\newcommand{\cmark}{\ding{51}}%
\newcommand{\xmark}{\ding{55}}%

\acrodef{TSE}{traditional search engine}
\acrodef{CIS}{conversational information seeking}
\acrodef{WISE}{wizard of search engine}
\acrodef{ID}{intent detection}
\acrodef{KE}{keyphrase extraction}
\acrodef{AP}{action prediction}
\acrodef{QS}{query selection}
\acrodef{PS}{passage selection}
\acrodef{RG}{response generation}
\acrodef{TDS}{task-oriented dialogue system}
\acrodef{KGC}{Knowledge-Grounded Conversation}

\makeatletter
\renewcommand\paragraph{\@startsection{paragraph}{4}{\parindent}%
  {0.25\baselineskip \@plus 0\p@ \@minus 0\p@}%
  {-3.5\p@}%
  {\ACM@NRadjust{\@parfont\@adddotafter}}}
\makeatother  

\author{Pengjie Ren$^{1}$\qquad Zhongkun Liu$^{1}$\qquad Xiaomeng Song$^{1}$\qquad Hongtao Tian$^{1}$
}
\author{Zhumin Chen$^{1*}$\qquad Zhaochun Ren$^{1*}$\qquad Maarten de Rijke$^{2,3}$
}

\affiliation{
 \institution{\textsuperscript{\rm 1}Shandong University, Qingdao, China}
 \institution{\textsuperscript{\rm 2}University of Amsterdam, Amsterdam, The Netherlands}
 \institution{\textsuperscript{\rm 3}Ahold Delhaize, Zaandam, The Netherlands}
 \country{}
}
\email{{renpengjie, chenzhumin, zhaochun.ren}@sdu.edu.cn, {m.derijke}@uva.nl} 
 
\renewcommand{\shortauthors}{Ren et al.}
\thanks{$^*$Corresponding author.}

\def\authors{Pengjie Ren, Zhongkun Liu, Xiaomeng Song, Hongtao Tian, Zhumin Chen, Zhaochun Ren, and Maarten de Rijke}

\begin{document}
\fancyhead{}

\title{Wizard of Search Engine}
\subtitle{Access to Information through Conversations with Search Engines}

% !TEX root =  ../main.tex

\begin{abstract}
\Ac{CIS} is playing an increasingly important role in connecting people to information.
Due to a lack of suitable resources, previous studies on \ac{CIS} are limited to the study of conceptual frameworks, laboratory-based user studies, or a particular aspect of \ac{CIS} (e.g., asking clarifying questions).

In this work, we make three main contributions to facilitate research into \ac{CIS}:
\begin{enumerate*}
\item We formulate a pipeline for \ac{CIS} with six sub\-tasks: \acl{ID}, \acl{KE}, \acl{AP}, \acl{QS}, \acl{PS}, and \acl{RG}.
\item We release a benchmark dataset, called \acfi{WISE}, which allows for comprehensive and in-depth research on all aspects of \ac{CIS}.
\item We design a neural architecture capable of training and evaluating both jointly and separately on the six sub-tasks, and devise a pre-train/fine-tune learning scheme, that can reduce the requirements of \ac{WISE} in scale by making full use of available data.
\end{enumerate*}

We report useful characteristics of the \ac{CIS} task based on statistics of the  \ac{WISE} dataset.
We also show that our best performing model variant is able to achieve effective \ac{CIS}.
We release the dataset, code as well as evaluation scripts to facilitate future research by measuring further improvements in this important research direction.
\end{abstract}


 

\maketitle

\acresetall

% !TEX root =  ../main.tex

\section{Introduction}
Search engines have become the dominant way for people around the world to interact with information. 
Today, \acp{TSE} have settled on a query-SERP (Search Engine Result Page) paradigm, where a searcher issues a query in the form of keywords to express their information need, and the search engine responds with a SERP containing a ranked list of snippets.
Even though this has become the dominant paradigm for interactions with search engines,
it is confined to clear navigational or transactional intents where the user wants to reach a particular site~\cite{10.1145/792550.792552}, or relatively simple informational intents, where the user's questions can be answered with short text spans~\cite{choi-etal-2018-quac}.
However, it is not effective when the user has complex and composite intents, especially when their information need is unclear and/or exploratory~\cite{DBLP:series/synthesis/2009White}.

\Acfi{CIS} has emerged as a new paradigm for interactions with search engines~\cite{10.1145/3397271.3401397,10.1145/3397271.3401415,DBLP:journals/corr/abs-2005-08658}.
In contrast to the query-SERP paradigm, \ac{CIS} allows users to express their information need by directly conducting conversations with search engines.
In this way, \ac{CIS} systems can better capture a user's intent by taking advantage of the flexibility of mixed-initiative interactions, and provide useful information more directly by returning human-like responses.
\ac{CIS} is increasingly attracting attention.
Currently, research on \ac{CIS} is mostly focused on the following aspects: asking clarifying questions~\cite{Aliannejadi:2019:ACQ:3331184.3331265,Hamedwww2020}, conversational search~\cite{10.1145/3397271.3401206,10.1145/3397271.3401130}, conversational question answering~\cite{reddy-etal-2019-coqa,ren-2020-conversations-arxiv}, and conversational 
recommendation~\cite{10.1145/2939672.2939746,10.1145/3209978.3210002}.
A complete \ac{CIS} system should be a mixture of these aspects and go far beyond these~\citep{gao-2021-advances-arxiv}.
Other aspects, e.g., chitchat~\cite{wu-yan-2018-deep}, user intent understanding~\cite{10.1145/3295750.3298924}, feedback analysis~\cite{10.1145/3308558.3313661}, conversation management, and so on~\cite{TRIPPAS2020102162}, are still not well studied in the context of \ac{CIS} scenario.
Since no complete \ac{CIS} models capable of handling all these aspects have been developed so far, there is a lack of comprehensive analysis on the performance of those aspects when achieved and/or evaluated simultaneously.
This is largely because there is no practical formalism or suitable resource for \ac{CIS}.
As a result, related research is restricted to user studies~\cite{Vtyurina:2017:ECS:3027063.3053175,10.1145/3176349.3176387} and/or theoretical/conceptual \ac{CIS} frameworks~\cite{10.1145/3020165.3020183,azzopardi2018conceptualizing}.

\begin{figure*}[htb]
\centering
\includegraphics[width=0.86\textwidth]{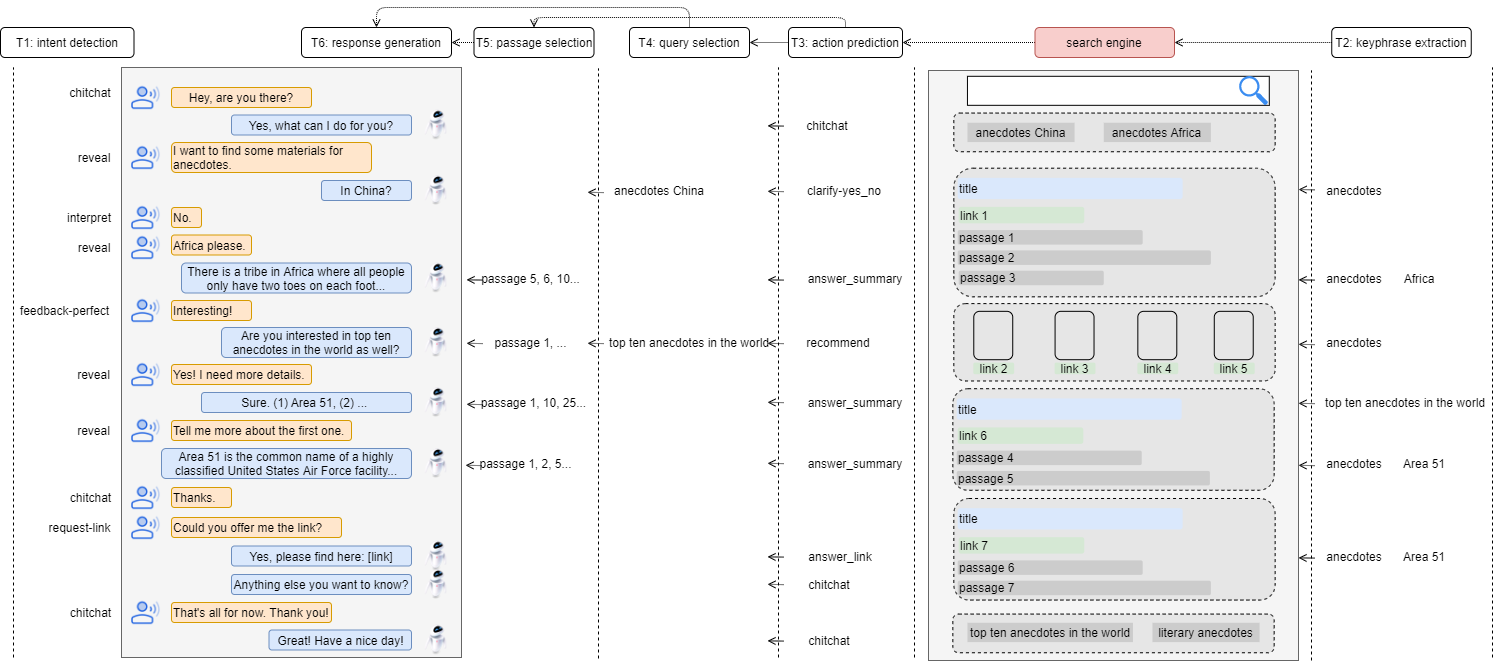}
\caption{Pipeline formalism for \acf{CIS}.}
\label{figure_task}
\end{figure*}

We contribute to \ac{CIS} by pursuing three goals in this work: a pipeline, a dataset, and a model.
Let us expand on each of these.
\begin{enumerate}[nosep,leftmargin=*]
\item We formulate a pipeline for \ac{CIS} consisting of six sub-tasks: \acf{ID}, \acf{KE}, \acf{AP}, \acf{QS}, \acf{PS}, and \acf{RG}, as shown in Figure~\ref{figure_task}. 
\ac{ID} identifies the general intent behind the current user utterance, e.g., greetings (``chitchat'') or revealing information need (``reveal'').
Then, \ac{KE} extracts keyphrases from previous conversations to represent the fine-grained information need behind the current user utterance (if any), which can be used to retrieve candidate queries (i.e., suggested queries, related queries) and candidate documents/passages.
\ac{AP} is used for conversation management as it decides the system action at each turn.
\ac{QS} and \ac{PS} select the relevant queries and passages from which \ac{RG} can distill useful information (if any) to generate conversational responses.
We make a close connection to \acp{TSE} so as to retain its characteristics as much as possible (e.g., candidate queries and passage retrieval) while introducing new characteristics of \ac{CIS} (e.g., conversation management, conversational response generation).
\item We build a dataset of human-human \ac{CIS} conversations in a wizard-of-oz fashion, called \acfi{WISE}, where two workers play the role of seeker and intermediary, respectively.
We first collect search sessions from a commercial search engine, which cover a diverse range of information needs.
Then, we write guidelines for the seekers by inferring the intents behind the keyword queries in the search sessions, e.g., ``[anecdotes, anecdotes in the world, Tolstoy, Tolstoy Wikipedia] $\rightarrow$ You want to find some materials about anecdotes. Describe the features of anecdotes you are interested in. Discover the introduction of the anecdotes and ask for links.''
The seekers can better understand the information need when provided with guidelines than keyword queries from search sessions.
We make sure the guidelines cover diverse search scenarios, e.g., when the users have specific, unclear, or multiple intents.
Finally, we build a conversational interface for the workers to conduct conversations.
The interface for the intermediaries is backed by a commercial search engine.
We record all the operations the workers take, which can be used for supervised learning or direct evaluation for all six sub-tasks.
\item We propose a modularized end-to-end neural architecture, where each module corresponds to one of the six sub-tasks listed above.
Thus, we have a natural way to train and evaluate the model, both jointly and separately, by either forwarding the ground truth or the prediction to downstream modules.
We also devise a pre-train/fine-tune learning scheme to make full use of available data and reduce the requirements of supervised learning.
We use four related tasks/datasets to guarantee that all the modules can be pre-trained.
\end{enumerate}

\noindent%
We conduct analyses to determine several useful characteristics of \ac{CIS} as reflected in the \ac{WISE} dataset.
We also carry out experiments to analyze the performance of different model variants on the six sub-tasks, as well as the effectiveness of the pre-training tasks.
We show that our best performing model variant is able to achieve effective \ac{CIS}.
To facilitate experimental reproducibility, we share the dataset, code, and evaluation scripts.
% We will also set up a leaderboard to advance future research by providing a fair platform to measure further improvements in \ac{CIS}.

% \begin{enumerate*}
% \item User studies. For example, \citet{Vtyurina:2017:ECS:3027063.3053175} conduct a user study and conclude that existing conversational assistants, e.g., Amazon Alexa, Apple Siri, and Microsoft Cortana, are not yet up to task, i.e., they cannot be effectively used for complex \ac{CIS} tasks.
% \item Theoretical/conceptual frameworks. For example, \citet{10.1145/3020165.3020183} present a formalism which describes how the \ac{CIS} system could be implemented, and makes the action space of a \ac{CIS} system explicit. 
% \item Clarifying questions. For example, \citet{Aliannejadi:2019:ACQ:3331184.3331265} formulate the task of asking clarifying questions in \ac{CIS} and \citet{Hamedwww2020} propose to generate clarifying questions for information retrieval.
% \item Conversational question answering. For example, \citet{10.1145/3397271.3401206} and \citet{ren-2020-conversations-arxiv} study how to retrieve relevant SERPs based on which to generate conversational responses for \ac{CIS}.
% \end{enumerate*}

%  Information seeking
% actions were not covered in depth but instead broad categories such as “answer” or “info-request” are presented. Recent work has
% made initial steps towards understanding such actions (Radlinski & Craswell, 2017; Trippas, Spina, Cavedon, & Sanderson, 2017b,
% Trippas, 2019), however, no complete set has been developed so far

% !TEX root =  ../main.tex

\section{Tasks}

Table~\ref{table_symbol} summarizes the notation used in the paper.

\begin{table}
\centering
\caption{Summary of main notation used in the paper.}
\label{table_symbol}
\begin{tabular}{lp{7cm}} 
\toprule
$X^{\tau}$ & User utterance at turn $\tau$. $X^{\tau}_{t}$ represents the token at position $t$.                                                                                       \\
$Y^{\tau}$ & System utterance at turn $\tau$. $Y^{\tau}_{t}$ represents the token at position $t$.                                                                                     \\
$C$                            & A conversation. $C^{:\tau}={\ldots, X^{\tau-1}, Y^{\tau-1}}$ represents the utterances until turn $\tau-1$.  \\
$d$                            & A passage $d \in \mathbb{D}$. $D^{\tau}$ represents the set of candidate passages at turn $\tau$.                                   \\
$q$                            & A query $q \in \mathbb{Q}$. $Q^{\tau}$ represents the set of candidate queries at turn $\tau$.                                      \\
$i$                            & A user intent $i \in \mathbb{I}$.                                                                                                                                          \\
$a$                            & A system action $a \in \mathbb{A}$.                                                                                                                                        \\
$k$                            & A keyphrase. $K^{\tau}$ is the set of keyphrases for $C^{:\tau}+X^{\tau}$.                                                                                                                                                                                                \\
$h$                            & A hidden representation $h \in H$. $H$ is a sequence.
% Superscript (if any) represents its type or conversation turn. Subscript (if any) corresponds to time steps.                  
\\
\bottomrule
\end{tabular}
\end{table}

\subsection{Task 1: Intent detection (ID)}

The goal of this task is to identify the general intent of each user utterance.
There have been several similar tasks~\cite{10.1145/3295750.3298924}.
However, the proposed user intent taxonomies are either too general or not well mapped to user behavior in \acp{TSE}.
Considering previous work~\cite{10.1145/3020165.3020183,azzopardi2018conceptualizing} while making a connection to \acp{TSE}, we define a new user intent taxonomy, as shown in Table~\ref{table_user_intent}.
Most intents can map to the corresponding user behavior or operation in \acp{TSE}, except for ``chitchat'' and ``request-rephrase'', because \acp{TSE} do not have conversations and the outputs are always organized as SERPs.
The algorithmic problem of \ac{ID} in \ac{CIS} is defined as learning a mapping: $\{C^{:\tau}, X^{\tau}\} \rightarrow i$.
That is, given the context $C^{:\tau}$ and the current user utterance $X^{\tau}$, the goal is to predict the user intent $i$ of for the utterance $X^{\tau}$.

\begin{table*}
\centering
\caption{User intent for the seeker.}
\label{table_user_intent}
\begin{tabular}{lp{5cm}p{6.5cm}p{3cm}}
\toprule
\multicolumn{1}{l}{\textbf{Intent}}               & \textbf{Explanation}                                                                                    & \textbf{Example}                                                                                                                                                 & \textbf{\ac{TSE} operations}                         \\
\midrule
\multirow{1}{*}[-0.2cm]{reveal}               & Reveal a new intent, or refine an old intent proactively.                                                   & \multirow{1}{*}[-0cm]{\begin{tabular}[c]{@{}l@{}}User: I want to see a movie. (reveal)\\User: Can you tell me more about it? (reveal)\end{tabular}}                                     & \multirow{1}{*}[-0.2cm]{Issue a new query.}                                   \\
\midrule
\multirow{1}{*}[-0.4cm]{revise}          & Revise an intent proactively when there is wrong expression, e.g., grammatical issues, unclear expression. & \multirow{1}{*}[-0.3cm]{\begin{tabular}[c]{@{}l@{}}User: Tell me some non-diary milks.\\User: I mean dairy not diary. (revise)\end{tabular}}                                               & \multirow{1}{*}[-0.4cm]{Revise the query.}                                    \\
\midrule
\multirow{1}{*}[-0.3cm]{interpret}       & Interpret or refine an intent by answering a clarification question from the system.                      & \multirow{1}{*}[-0cm]{\begin{tabular}[c]{@{}l@{}}User: Do you know The Avengers?\\System: Do you mean the movie, novel or game?\\User: The movie (interpret)\end{tabular}}                  & \multirow{1}{*}[-0.3cm]{Select 
suggested queries.}         \\
\midrule
\multirow{1}{*}[-0.2cm]{request-rephrase} &  Request the system to rephrase the response if it is not understandable.                                & \multirow{1}{*}[-0.2cm]{Sorry, I didn’t get it. (request-rephrase)}  &         --                           \\
\midrule
\multirow{1}{*}[-0.1cm]{chitchat}           & Greetings or other utterances that are not related to the information need.                              & \multirow{1}{*}[-0cm]{\begin{tabular}[c]{@{}l@{}}I see. (chitchat)\\Are you there? (chitchat)\end{tabular}}                                              &       --
\\\bottomrule
\end{tabular}
\end{table*}

\subsection{Task 2: Keyphrase extraction (KE)}
Conversations usually contain more content than necessary in order to ensure the integrity of grammar and semantics, etc.
Ellipsis and anaphora are common in conversations, so the current user utterance itself is usually not semantically complete.
This is especially true as conversations go on and when the users are not familiar with the topic, or unclear about the information goals themselves.
It is challenging to fetch documents that are relevant to the intent in the current user utterance by simply issuing the current user utterance to the \ac{TSE}.
To this end, similar to \cite{10.1145/3397271.3401130}, we define \ac{KE} as a task to extract keyphrases from the current user utterance as well as the previous conversational context, which can help \acp{TSE} to retrieve relevant documents for the current user utterance.
Formally, \ac{KE} is defined as learning a mapping: $\{C^{:\tau}, X^{\tau}\} \rightarrow K^{\tau}$.

\subsection{Task 3: Action prediction (AP)}

To connect users to the right information with maximal utility, a good \ac{CIS} system should be able to take appropriate actions at the right time, e.g., helping the users to clarify their intents whenever unclear, and providing information in the right form by tracking user expectations.
\ac{AP} has been well studied in \acp{TDS} in a closed-domain setting (a.k.a. dialogue policy)~\cite{peng-etal-2017-composite}.
However, \ac{AP} is not yet well addressed in \ac{CIS}.
Although there have been some investigations about what system actions are necessary in \ac{CIS}, to the best of our knowledge, it has not yet been put into practice~\cite{10.1145/3020165.3020183,azzopardi2018conceptualizing}.
By recalling the operations in \acp{TSE}, we introduce a system action taxonomy in Table~\ref{table_system_action}.
The \ac{AP} task is formulated as learning a mapping: $\{C^{:\tau}, X^{\tau}, Q^{\tau}, D^{\tau}\} \rightarrow a$.
Note that this task also takes the candidate queries $Q^{\tau}$ and candidate passages $D^{\tau}$ fetched from \acp{TSE} as input, as they will influence the action the \ac{CIS} system should take.
For example, to decide whether ``no-answer'' is the action to be taken, the \ac{CIS} system must check whether the answer is available in $Q^{\tau}$ and/or $D^{\tau}$.
% Note that the differences of ``answer'' and ``recommend'' are two-fold: ``answer'' is usually more objective (opinion answer is an exception), while ``recommend'' is more subjective; the response of ``recommend'' is usually a list of items (e.g., products).

\begin{table*}
\centering
\caption{System actions for the intermediary.}
\label{table_system_action}
\begin{tabular}{lp{1cm}p{5cm}p{5.7cm}p{2.5cm}} 
\toprule
\multicolumn{2}{l}{\textbf{Action} } & \textbf{Explanation}                                                                                                                                                                                                 & \textbf{Example}                                                                                                                    & \textbf{TSE operations}       \\
\midrule
\multirow{3}{*}[-0.5cm]{clarify} & yes-no    & \multirow{3}{*}[-0.5cm]{\begin{tabular}[c]{@{}l@{}}Ask
questions to clarify user intent\\when it is unclear or exploratory.\end{tabular}}                                                                                                                                             & Do
you
want
to the plot? (clarify-yes-no)                                                                                           & \multirow{3}{*}[-0.5cm]{Suggest
queries.}  \\
\cmidrule{2-2}\cmidrule{4-4}
                         & \multirow{1}{*}[-0.2cm]{choice}     &                                                                                                                                                                                                                       & Do
you want to know its plot, cast or director? (clarify-choice)                                                                     &                                    \\
\cmidrule{2-2}\cmidrule{4-4}
                         & \multirow{1}{*}[-0.2cm]{open}       &                                                                                                                                                                                                                       & What
information do you want to know? (clarify-open)                                                                                &                                    \\
\midrule
\multirow{3}{*}[-1cm]{answer-type}  & \multirow{1}{*}[-0.3cm]{opinion}    & Give advice, ideas, suggestions, or instructions. The response is more subjective.
 & \multirow{1}{*}[-0.3cm]{I recommend xxx, because … (answer-opinion)}                                                               & \multirow{7}{*}[-2cm]{Provide
results.}  \\
\cmidrule{2-4}
 & \multirow{1}{*}[-0.2cm]{fact}       & Give a single, unambiguous answer. The response is objective and certain. & \multirow{1}{*}[-0.3cm]{Her birthday is xxx. (answer-fact)}    &                                    \\
\cmidrule{2-4}
& \multirow{1}{*}[-0.2cm]{open} &   Give an answer to an open-ended question, or one with unconstrained depth. The response is objective but may be different depending on the perspectives.   & \multirow{1}{*}[-0.3cm]{\begin{tabular}[c]{@{}l@{}}One of the reasons of the earthquake is that…\\(answer-open)\end{tabular}}
  &  \\
\cmidrule{1-4}
\multirow{3}{*}[-0.8cm]{answer-form}  & \multirow{1}{*}[-0.1cm]{free-text}    & \multirow{4}{*}[-0.6cm]{\begin{tabular}[c]{@{}l@{}}Answer the user intent by providing\\information in the right form or when\\being asked to answer in a particular\\form.\end{tabular}} & The disadvantages of Laminate Flooring are that …… (answer\_free\_text)&\\
\cmidrule{2-2}\cmidrule{4-4}
& \multirow{1}{*}[0cm]{list}       &  & Area 51. … (answer\_list) &  \\
\cmidrule{2-2}\cmidrule{4-4}
& \multirow{1}{*}[0cm]{steps}       &  & 1. Click on … 2. (answer\_steps) &  \\
\cmidrule{2-2}\cmidrule{4-4}
& \multirow{1}{*}[-0.1cm]{link} &    & You can find the video here: [link]. (answer\_link)       &   \\
\midrule
\multirow{1}{*}[-0.2cm]{no-answer}                 &            & If there is no relevant information found, notice the user.   & \multirow{1}{*}[-0cm]{\begin{tabular}[c]{@{}l@{}}Sorry, I cannot find any relevant information.\\(no-answer)\end{tabular}} &       \multirow{1}{*}[-0.2cm]{No answer found.}                             \\
\midrule
\multirow{2}{*}[-0.1cm]{request-rephrase} &   & Ask
the user to rephrase its question if it is unclear.                       & I
didn’t really get what you mean. (request-rephrase)                                                                               &                     --               \\
\midrule
\multirow{1}{*}[-0.2cm]{chitchat}                 &            & Greetings
or other content that are not related to the information need.                                                                                                                             & \multirow{1}{*}[-0cm]{\begin{tabular}[c]{@{}l@{}}Hi. (chitchat)\\Yes, I am ready to answer your questions. (chitchat)\end{tabular}} &             --                       \\
\bottomrule
\end{tabular}
\end{table*}

\subsection{Task 4: Query selection (QS)}

In \acp{TSE}, a list of suggested or related queries is usually presented to users to help them clarify intents or find similar intents.
In \ac{CIS}, we cannot always simply list the suggested or related queries to users, as this is not natural in conversations.
Previous studies cast this problem as a problem of asking clarifying questions, and model it as either selecting a question from a question pool prepared in advance~\cite{Aliannejadi:2019:ACQ:3331184.3331265} or generating a clarifying question for a given keyword query~\cite{Hamedwww2020}.
However, neither option fits in our formulation.
Besides, how to recommend similar search intents rather than asking clarifying questions is neglected.
We introduce the \ac{QS} task as a step in the \ac{CIS} pipeline.
Depending on the predicted actions from the upstream \ac{AP} task, \ac{QS} aims to select queries fetched from \acp{TSE} that can be used by the downstream task to generate clarifying questions or recommend similar search intents.
Formally, \ac{QS} is defined as learning a mapping: $\{C^{:\tau}, X^{\tau}, Q^{\tau}, (i, a)\} \rightarrow Q^{\tau}_{s}$, where $Q^{\tau}$ is set of suggested or related queries fetched from \acp{TSE} with the keyphrases from \ac{KE} as queries, and $i$ and $a$ are the user intent and system action from \ac{ID} and \ac{AP}, respectively.

\subsection{Task 5: Passage selection (PS)}
Passage or document ranking has long been studied in \acp{TSE}~\cite{10.1145/3331184.3331233}.
Recently, this has also been investigated in a conversational context~\cite{10.1145/3397271.3401206}.
The situation is a bit different for \ac{PS} in \ac{CIS} where the context is not only about the proactive questions from users but also the clarifying questions or recommendations from the systems.
And the goal is not always to find the relevant passages for the user questions but also non-relevant passages that contain related topics in order to do clarification or recommendation sometimes~\cite{DBLP:series/synthesis/2009White}.
Given the same input as in \ac{QS}, the \ac{PS} task is to select passages based on which to generate system responses by the downstream task: $\{C^{:\tau}, X^{\tau}, D^{\tau}, (i, a)\} \rightarrow D^{\tau}_{s}$, where $D^{\tau}$ is a set of passages fetched from a \ac{TSE} with the same keyphrases from \ac{KE} as queries.

\subsection{Task 6: Response generation (RG)}

\ac{RG} is a fundamental component for many NLP tasks, e.g., chitchat \cite{ren-2020-thinking}, \ac{TDS}~\cite{pei-2020-retrospective}, conversational question answering~\cite{baheti-etal-2020-fluent}.
The goals or formulations of \ac{RG} vary across tasks.
The role of \ac{RG} in our formulation of \ac{CIS} is to translate system actions and predictions of the above tasks into natural language responses.
For example, if the system action is ``clarify'', the \ac{RG} part needs to generate a clarifying question by referring to the selected queries.
% In a more ideal scenario, the \ac{CIS} system should be able to generate responses in diverse forms, e.g., images, tables, etc.
In this work, we only focus on free text responses.
The formulation of \ac{RG} is defined as a learning problem: $\{C^{:\tau}, X^{\tau}, Q^{\tau}, D^{\tau}, (i, a, Q^{\tau}_{s}, D^{\tau}_{s})\} \rightarrow Y^{\tau}$.

% !TEX root =  ../main.tex

\section{Dataset}

\begin{figure}[t]
\centering
\includegraphics[width=0.9\columnwidth]{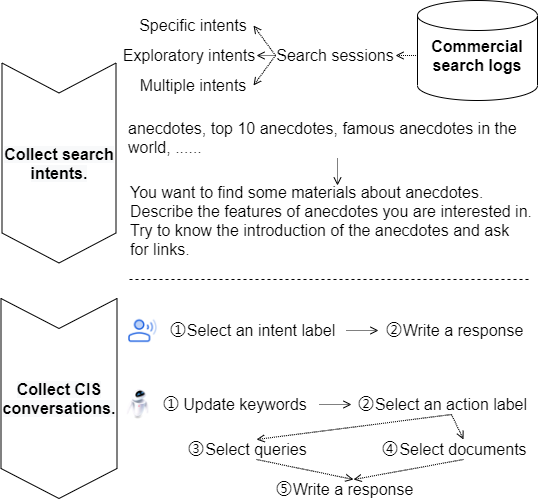}
\caption{Pipeline for data construction.}
\label{figure_data}
\end{figure}

The \ac{WISE} dataset is built based on the following setting: two participants engage in a \ac{CIS} conversation, one plays the role of a searcher while the other plays the role of a knowledgeable expert (referred to as wizard).
At each stage of the conversation, the searcher talks to the wizard freely. 
Their goal is to follow the general instruction about a chosen search intent, go into depth about it, and try to get all information that has been mentioned in the instruction.
The wizard helps the searcher to find the information by interacting with a search engine.
Their goal is to help the searcher (1) clarify their search intents whenever unclear, 
(2) answer the searcher's questions by finding and summarizing the relevant information in search results, and (3) recommend information that has not been asked but is related to the searcher's search intents or other related topics the searcher might be interested in.
There are two steps in building the \ac{WISE} dataset, as shown in Figure~\ref{figure_data}: collecting search intents, and collecting \ac{CIS} conversations.

\subsection{Collecting search intents}
First, we collect a set of 1,196 search intents from a search log of a commercial search engine.
Each search intent is based on a specific search session.
For a given search session, e.g., ``[anecdotes, anecdotes in the world, Tolstoy, Tolstoy Wikipedia, Tolstoy movie]'', the workers are asked to infer the search intent behind it using their imagination, and write down a description in one of the following forms: (1) specific intent, e.g., ``You are interested in anecdotes about famous writers in the world like Tolstoy. Try to learn about his/her basic information, life story and representative works, and so on.'' (2) exploratory intent, e.g., ``You want to find some material about anecdotes. Describe the features of anecdotes you are interested in. Try to learn about the introduction of the anecdotes and ask for links.'' (3) multiple intents, e.g., ``(1) or (2) + You also want to find a movie that is based on his/her works. Ask for information about the director, plot, links and so on.''

\subsection{Collecting CIS conversations}

The conversation flows as follows.
\begin{enumerate*}
\item The searcher randomly picks a search intent and starts by asking questions directly or by uttering a greeting. When sending a message, the searcher is required to select a label from Table~\ref{table_user_intent} that can best describe the intent of the message.
\item Whenever the wizard receives the message from the searcher, s/he needs to extract keyphrases from the conversation history that are used to fetch results from a search engine. Then s/he has to select a label from Table~\ref{table_system_action} that reflects the action s/he is going to take. After that, s/he needs to select the relevant queries and/or documents, based on which to formulate the responses.
\item The conversation repeats until the searcher ends the chat (after a minimum of 7 turns each).
\end{enumerate*}
For each turn, both the searcher and wizard can send multiple messages at once by clicking the ``send another message'' option.
After collecting data in this way, the goal is to then replace the wizard with a learned \ac{CIS} system that will speak to a human searcher instead.

\subsection{Measures/Disclaimers for data quality}

\subsubsection{Measures}
We took several measures to guarantee the data quality.
First, we provided detailed documentation on  the task, the definition of the labels, the steps involved in creating the data, and the instruction of using the interfaces.
Second, we recorded a video to demonstrate the whole process and enhance the considerations to take into account.
Third, we also maintained a list of positive and negative examples from us as well as the workers as a reference for the other workers.
Fourth, after obtaining the data, we manually went over it and corrected the following issues: grammatical issues, mis-labels, long responses without summarization.

\subsubsection{Disclaimers} 
In all the provided instruction materials, we described the purpose of this data construction effort and pointed out that the data will only be used for research.
We did not record any information about the workers and warned the workers not to divulge any of their private information during conversations.
We filtered out all adult content and/or offensive content that might raise ethical issues when collecting the search intents, and asked the workers to be careful about such content in conversations too.

\subsection{Statistics}
It took 24 workers 3 months to build the dataset.
The final dataset we collected consists of 1,905 conversations with 37,956 utterances spread over 1,196 search intents.
In total, the dataset contains 12 different Intents and 23 different Actions, covering a variety of conversation topics such as literature, entertainment, art, etc.
Each conversation contains 7 to 42 utterances. The average number of utterances per conversation is 19.9, and the average number of turns per conversation is 9.2. Each utterance has an average of 27.3 words.
We divide the data into 705 conversations for training, 200 conversations for validation, and 1,000 for testing.
The test set is split into two subsets, test(seen) and test(unseen). 
The test(seen) dataset contains 442 overlapping search intents with the training set with new conversations. 
Test(unseen) consists of 500 search intents never seen before in train or validation. 
The overall data statistics can be found in Table~\ref{table_dataset_statistics}.

% \begin{figure}[t]
% \centering
% \includegraphics[width=0.9\columnwidth]{figures/train_test_valid.png}
% \caption{Distribution of actions and intents in partitioned datasets.}
% \label{figure_split_data}
% \end{figure}

\begin{table}[t]
\centering
\caption{Datasets statistics.}
\label{table_dataset_statistics}
\begin{tabular}{l r r rr}
\toprule
\multirow{2}{*}{} & \multirow{2}{*}{train} & \multirow{2}{*}{valid} & \multicolumn{2}{c}{test}                       \\ \cline{4-5}
                  &                        &                        & \multicolumn{1}{c}{test(seen)} & test(unseen) \\ \midrule
conversations     & 705                    & 200                    & 500        & 500          \\
turns             & 6,099                   & 1,925                   & 4,860       & 4,626         \\
utterances        & 13,569                  & 4,125                   & 10,410      & 9,852         \\
avg turns         & 8.65                   & 9.63                   & 9.72       & 9.25         \\
avg utterances    & 19.25                  & 20.63                  & 20.82      & 19.70        \\
avg words         & 27.06                  & 29.3                   & 25.94      & 28.3         \\ 
\bottomrule
\end{tabular}
\end{table}

% The distribution of intent and action labels over different turns are shown in Figure~\todo{xxx}.

% !TEX root =  ../main.tex

\section{Methodology}

We propose a model to replace the wizard in data construction and devise a pre-train/fine-tune learning scheme to train the model.

\subsection{Model}
\begin{figure}[ht]
\centering
\includegraphics[width=0.9\columnwidth]{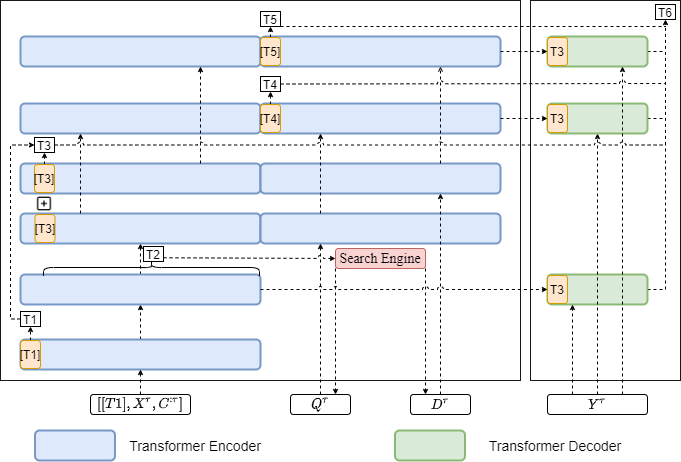}
\caption{Model architecture.}
\label{figure_model}
\end{figure}

As shown in Figure~\ref{figure_model},  we propose an architecture that relies mainly on transformer encoders and decoders.

\subsubsection{Transformer encoder} 
The transformer encoder consists of two layers: a self-attention layer and
a position-wise feed-forward layer. 
The self-attention layer adopts the multi-head attention mechanism~\cite{10.5555/3295222.3295349} to identify useful information in the input sequence. 
The feed-forward layer consists of two linear transformations with a ReLU activation in between~\cite{DBLP:journals/corr/abs-1803-08375}.
Before each layer, layer
normalization is applied~\cite{ba2016layer}.
After each layer, a residual connection is applied.
For any given sequence $S$, the transformer encoder outputs a sequence of hidden representations $H \in \mathbb{R}^{l_S \times dim}$ corresponding to $S$: $H = \operatorname{TEncoder}(S)$, where $l_S$ is the sequence length, $dim$ is the hidden size, and $TEncoder$ is short for transformer encoder.

\subsubsection{Transformer decoder} 
The transformer decoder consists of three layers: an output-attention layer, an input-attention layer and a position-wise feed-forward layer.
They all use the same implementation as in the transformer encoder.
The difference is that the output-attention and input-attention layers try to identify useful information in the output sequence and input sequence, respectively.
For the output sequence, a look-ahead mask is used to mask the future positions so as to prevent information from future positions from affecting current positions~\cite{10.5555/3295222.3295349}.
Again, before and after each layer, a layer
normalization and a residual connection are applied, respectively.
For any given output and input sequences $S_{O}$ and $S_{I}$, the transformer decoder outputs a sequence of hidden representations $H \in \mathbb{R}^{l_O \times dim}$ corresponding to $S_{O}$ and the input-attention weights $W \in \mathbb{R}^{l_O \times l_I}$ corresponding to $S_{I}$: $H, W = \operatorname{TDecoder}(S_{O}, S_{I})$, where $l_O$ and $l_I$ are the output and input sequence lengths, respectively.
$\operatorname{TDecoder}$ is short for transformer decoder.

\subsubsection{Task 1: \Acf{ID}.}
We concatenate $C^{:\tau}$ and $X^{\tau}$ in reverse order of utterance turns, and put a special token ``[T1]'' at the beginning to get the input of Task 1: $S_{T1} = [[T1], X^{\tau}, C^{:\tau}]$.
The sequence goes to a token and positional embedding layer to get $E_{C}$.
Then, we input the embedding sequence into a stack of transformer encoders to get hidden representations for all tokens: $H_{T1} = \operatorname{TEncoder}(E_{C})$.
Finally, we get the hidden representation w.r.t.\ ``[T1]'' and use a linear classifier with softmax to do \ac{ID}. 

\subsubsection{Task 2: \Acf{KE}.}
We model \ac{KE} as a binary classification on each token of $C^{:\tau}$ and $X^{\tau}$.
To do so, the same stack of transformer encoders as for \ac{ID} is used to get the hidden representations for \ac{KE} based on the representations from \ac{ID}: $H_{T2} = \operatorname{TEncoder}(H_{T1})$.
A linear classifier with sigmoid is used to predict whether each token belongs to a keyphrase or not.

\subsubsection{Task 3: \Acf{AP}.}
To predict the system action, we also need to consider the query and passage candidates $Q^{\tau}$ and $D^{\tau}$.
For each $q$ or $d$, we first input it into a token and positional embedding layer to get $E_{q}$ or $E_{d}$.
We then concatenate it with the hidden representations from \ac{KE} and fuse the context and query/passage information with the same stack of transformer encoders as for \ac{ID}: $H^q_{T3} = \operatorname{TEncoder}([H_{T2}, E_{q}])$ and $H^d_{T3} = \operatorname{TEncoder}([H_{T2}, E_{d}])$.
Note that another special token ``[T3]'' is used (see Figure~\ref{figure_model}).
A hidden representation corresponding to ``[T3]'' is obtained for each $q$ or $d$.
We combine them with a max pooling to get a single hidden representation, on top of which a linear classifier with softmax is applied to do \ac{AP}.

\subsubsection{Task 4: \Acf{QS}}
For \ac{QS}, a similar architecture as \ac{AP} is adopted.
The same stack of transformer encoders is used to get hidden representations for \ac{QS} based on the hidden representations $H^q_{T3}$ from \ac{AP}: $H_{T4} = \operatorname{TEncoder}(H^q_{T3})$.
We get the hidden representation corresponding to ``[T4]'', on top of which a binary classifier with sigmoid is used to predict whether a query $q$ should be selected or not.

\subsubsection{Task 5: \Acf{PS}}
Similar to \ac{QS}, \ac{PS} uses the same stack of transformer encoders to get hidden representations based on the hidden representations $H^d_{T3}$ from \ac{AP}: $H_{T5} = \operatorname{TEncoder}(H^d_{T3})$.
The hidden representation corresponding to ``[T5]'' is used to predict whether a passage $d$ should be selected or not using a binary classifier with sigmoid.

\subsubsection{Task 6: \Acf{RG}}
We use a stack of transformer decoders to do \ac{RG}.
The output sequence in the transformer decoder is the ground truth response $Y^{\tau}$ during training or the generated response $\hat{Y}^{\tau}$ during inference.
Note that we put the system action at the beginning to indicate the type of the response.
We get the fused hidden representations by taking the context, query, and passage hidden representations from \ac{KE}, \ac{QS}, and \ac{PS} as the input sequence in the transformer decoder, respectively.
Specifically, $H^{C}_{T6}, W^{C}_{T6} = \operatorname{TDecoder}(E_{Y}, H_{T2})$; $H^{q}_{T6}, W^{q}_{T6}  = \operatorname{TDecoder}(E_{Y}, H_{T4})$; $H^{d}_{T6}, W^{d}_{T6}  = \operatorname{TDecoder}(E_{Y}, H_{T5})$.
$H^{C}_{T6}$, $H^{q}_{T6}$ and $H^{d}_{T6}$ are fused again with max pooling.
Then a linear classifier with softmax is applied on top to predict the probability of the token at each time step $P_{g}(Y^{\tau}_{t})$.
We also extend the pointer mechanism~\cite{10.5555/2969442.2969540} to generate a token by copying a token in the context, query or passage~\cite{see-etal-2017-get}.
Pointer mechanisms are commonly used in natural language generation tasks.
The copying probability is modeled as $P_{c}(Y^{\tau}_{t}) = \max (W^{C}_{T6}(Y^{\tau}_{t}), P(q)W^{q}_{T6}(Y^{\tau}_{t}), P(d)W^{d}_{T6}(Y^{\tau}_{t}))$, where $W^{C}_{T6}(Y^{\tau}_{t})$ is the sum of attention weights for $Y^{\tau}_{t}$ in the context.
A similar definition applies to $W^{q}_{T6}(Y^{\tau}_{t})$ and $W^{d}_{T6}(Y^{\tau}_{t}))$.
$P(q)$ and $P(d)$ are the selection probability from \ac{QS} and \ac{PS}, respectively.
The final probability $P(Y^{\tau}_{t})$ is a linear combination of $P_{g}(Y^{\tau}_{t})$ and $P_{c}(Y^{\tau}_{t})$.

\subsection{Learning}
We use binary cross-entropy loss for \ac{KE}, \ac{QS} and \ac{PS} and cross-entropy loss for \ac{ID}, \ac{AP} and \ac{RG}.
To solve complex tasks with a deep model, a large volume of data is needed.
We get around this problem by first pre-training the model on four datasets from related tasks, as shown in Table~\ref{table_pre-training}.
\begin{table}[h]
\centering
\caption{Pre-training datasets.}
\label{table_pre-training}
\setlength{\tabcolsep}{1pt}
\begin{tabular}{l @{} c c c c c c}
\toprule
\textbf{Datasets}                            & \textbf{T1 (\ac{ID})}                      & \textbf{T2 (\ac{KE})}                                    & \textbf{T3 (\ac{AP})}                     & \textbf{T4 (\ac{QS})}                           & \textbf{T5 (\ac{PS})}                & \textbf{T6 (\ac{RG})}                  \\ \midrule
WebQA~\cite{DBLP:journals/corr/LiLHWCZX16}       &    \xmark                     &         \xmark                              &          \xmark              &            \xmark                  & \cmark & \cmark   \\
DuReader~\cite{he-etal-2018-dureader}  & \cmark &    \xmark                                   & \cmark &      \xmark                        & \cmark & \cmark \\ 
DuConv~\cite{wu-etal-2019-proactive}     &         \xmark                & \cmark                &              \xmark          & \cmark   &         \xmark          & \cmark \\
KdConv~\cite{zhou-etal-2020-kdconv}   &     \xmark                    & \cmark &      \xmark                  & \cmark &    \xmark               & \cmark \\ 
\bottomrule
\end{tabular}
\end{table}
WebQA is a dataset for single-turn question answering.
For each sample, it provides a simple question in natural language, a set of candidate passages, and an entity answer.
DuReader is for single-turn machine reading comprehension.
The questions are mostly complex and the answers are summaries of the relevant passages.
It also has question and answer types, which can be used for the pre-training of T1 and T3.
DuConv and KdConv are for knowledge grounded conversations.
DuConv has structured knowledge in triples while KdConv has free text sentences.
Since the sentences in KdConv are usually short, we use it to pre-train T4.
DuConv also has annotations of topic words in conversations which can be used to pre-train T2.
For KdConv, we pre-train T2 by predicting the overlapping tokens in conversational contexts and answers.
We do not pre-train T2 using WebQA or DuReader, because they are single-turn and the questions are very short usually.
After pre-training, we fine-tune the model on the \ac{WISE} data. 
 
% !TEX root =  ../main.tex

\section{Experimental Setup}
% We seek to answer the following questions in our experiments:
% \begin{enumerate*}[leftmargin=*,label=(RQ\arabic*)]
% \item What performance can we achieve on the \ac{WISE} dataset?
% %three tables: our method vs baseline, our method w.r.t. different intents, our method w.r.t. different actions
% \item How do the different subtasks contribute to the performance?
% %one table: ablation study
% \item What are the effects of different datasets for pre-training?
% %one table: performance w.r.t. different pre-training datasets
% % \item Is there any room for further improvement?
% \end{enumerate*}

% \subsection{Datasets}

% \subsection{Baselines}
% We compare \ac{WISE} with state-of-the-art \ac{KGC} method, TMemNet~\cite{conf/iclr/DinanRSFAW19}.
% It was first introduced for the knowledge grounded dialogue task.
% It combines a Memory Network and Transformer to do knowledge selection and response generation.

\paragraph{Evaluation metrics}
We adopt different evaluation metrics for different subtasks. 
We use BLEU-1~\cite{conf/acl/PapineniRWZ02} and ROUGE-L~\cite{lin-2004-rouge} for \acf{KE} and \acf{RG}, which are commonly used in NLG tasks, e.g., QA, MRC~\cite{journals/tacl/KociskySBDHMG18,conf/nips/NguyenRSGTMD16}.
We report macro Precision, Recall and F1 for \acf{ID}, \acf{AP}, \acf{QS} and \acf{PS}.

\paragraph{Implementation details}
For a fair comparison, we implement all models used in our experiments based on the same code framework.
% to ensure that they share the same code apart from the model itself.
We set the word embedding size and hidden size to 512. 
We use the same vocabulary for all methods and its size is 21,129. 
The learning rate was increased linearly from zero to $2.5 \times 10^{-4}$ in the first 6,000 steps and then annealed to 0 by using a cosine schedule.
We use gradient clipping with a maximum gradient norm of 1.
We use the Adam optimizer ($\alpha$ = 0.001, $\beta_1$ = 0.9, $\beta_2$= 0.999, and $\epsilon$ = $10^{-8}$).
We use 4 transformer layers for the encoder and 2 for the decoder, where the number of attention heads is 8.
In WISE, the parameters of the transformer encoder, transformer decoder, and word embedding for each component of our model are shared. 
% The baseline has the same parameters settings as the WISE. 
We train all models for 50 epochs, evaluate on a validation set for the best model in terms of BLEU-1 in \ac{RG} task. 
%The code and data are available online.\footnote{\url{https://github.com/PengjieRen/CaSE_WISE}.}

% !TEX root =  ../main.tex

\section{Result and Analysis}

\subsection{Performance analysis}
We report the results of all methods on \ac{WISE}; see Table~\ref{tab: result}.
We consider three settings:
\begin{enumerate*}
\item With pretraining~(i.e., WISE in Table~\ref{tab: result}).
\item Without pretraining~(i.e., WISE-pretrain in Table~\ref{tab: result}).
\item With pretraining and using the ground truth of \acs{ID}, \acs{KE}, \acs{AP}, \acs{QS} and \acs{PS}~(i.e., WISE+GT in Table~\ref{tab: result}).
\end{enumerate*}
We also report the performance of \ac{WISE} on the test~(unseen) and test~(seen) datasets, and on different actions. 
The results are shown in Table~\ref{tab: result_seen_unseen} and \ref{tab: DiffAction}.

\begin{table*}[ht]
\centering
\caption{Overall performance. Bold face indicates the best result in terms of corresponding metric. Significant improvements over the strongest \ac{WISE} results are marked with $^{*}$ (p-value < 0.01 with t-test).}
\label{tab: result} 
\setlength{\tabcolsep}{1.2mm}{
\begin{tabular}{lcccccccccccccccc}
\toprule
     & \multicolumn{3}{c}{ID~(\%)} & \multicolumn{2}{c}{KE~(\%)} & \multicolumn{3}{c}{AP~(\%)} & \multicolumn{3}{c}{QS~(\%)} & \multicolumn{3}{c}{PS~(\%)} & \multicolumn{2}{c}{RG~(\%)} \\
\cmidrule(lr){2-4} \cmidrule(lr){5-6} \cmidrule(lr){7-9} \cmidrule(lr){10-12} \cmidrule(lr){13-15} \cmidrule(lr){16-17}
     & P     & R     & F1     & BLEU      & ROUGE      & P     & R     & F1     & P     & R     & F1     & P     & R     & F1     & BLEU      & ROUGE      \\
\midrule
WISE-pretrain   &   \textbf{71.3}    &     12.5  &    11.7    &  0.0         &      0.0      &    12.7   &   4.5    &      1.8  &    0.0   &  0.0     &  0.0      &     0.0  & 0.0      &  0.0      &    1.0       &  1.0  \\ 
WISE &    45.2   &    \textbf{32.5}   &  \textbf{34.1}      &    \textbf{58.3}      & \textbf{59.0}    &   \textbf{18.8}    &  \textbf{20.6}      &    \textbf{17.8}   &  \textbf{26.9}     &  \textbf{16.1}      &    \textbf{11.5}   &   \textbf{35.8}    &  \textbf{31.5}      &    \textbf{24.9}       &  14.6 & 16.9  \\ 
WISE+GT     &    45.2   &    \textbf{32.5}   &  \textbf{34.1}      &    \textbf{58.3}      & \textbf{59.0}   &   \textbf{18.8}    &  \textbf{20.6}     &    \textbf{17.8}  &  \textbf{26.9}    &  \textbf{16.1}      &    \textbf{11.5}   &   \textbf{35.8}    &  \textbf{31.5}      &    \textbf{24.9}       &  \textbf{20.4}\rlap{$^{*}$} & \textbf{24.2}\rlap{$^{*}$}  \\  
\bottomrule
\end{tabular}

}
\end{table*}

\begin{table*}[ht]
\centering
\caption{Overall performance of WISE on the test(seen) and test(unseen) dataset.}
\label{tab: result_seen_unseen} 
\setlength{\tabcolsep}{1.2mm}{
\begin{tabular}{lcccccccccccccccc}
\toprule
     & \multicolumn{3}{c}{ID~(\%)} & \multicolumn{2}{c}{KE~(\%)} & \multicolumn{3}{c}{AP~(\%)} & \multicolumn{3}{c}{QS~(\%)} & \multicolumn{3}{c}{PS~(\%)} & \multicolumn{2}{c}{RG~(\%)} \\
\cmidrule(lr){2-4} \cmidrule(lr){5-6} \cmidrule(lr){7-9} \cmidrule(lr){10-12} \cmidrule(lr){13-15} \cmidrule(lr){16-17}
     & P     & R     & F1     & BLEU      & ROUGE      & P     & R     & F1     & P     & R     & F1     & P     & R     & F1     & BLEU      & ROUGE      \\
\midrule
test~(unseen)      &    38.4   &    28.5   &  29.3     &    \textbf{59.2}      & \textbf{59.0}    &   17.6    &  18.1      &    16.5   &  \textbf{28.8}    &  15.1      &    10.8   &   34.2    &  26.7      &    20.5       &  14.1 & 16.0  \\ 
test~(seen)   &    \textbf{48.3}   &    \textbf{36.1}  &    \textbf{37.4}    &  57.3        &      58.4      &    \textbf{19.9}   &   \textbf{24.2}    &      \textbf{19.0}  &    25.4   &  \textbf{21.5}    &  \textbf{15.4}     &    \textbf{36.8}  &  \textbf{37.6}      &  \textbf{29.6}     &    \textbf{14.6}     &  \textbf{17.4}  \\ 
test &    45.2   &    32.5   &  34.1      &    58.3       & \textbf{59.0}    &   18.8    &  20.6      &    17.8   &  26.9     &  16.1      &    11.5   &   35.8    &  31.5      &    24.9       &  \textbf{14.6} & 16.9  \\ 

\bottomrule
\end{tabular}

}
\end{table*}

\begin{table*}[]
\centering
\caption{Overall performance of \acs{WISE} for different actions.}
\label{tab: DiffAction}
\begin{tabular}{l l ccccccccccccc}
\toprule
\multicolumn{2}{c}{\multirow{2}{*}{}} & \multicolumn{2}{c}{KE~(\%)} & \multicolumn{3}{c}{AP~(\%)} & \multicolumn{3}{c}{QS~(\%)} & \multicolumn{3}{c}{PS~(\%)} & \multicolumn{2}{c}{RG~(\%)} \\
\cmidrule(lr){3-4} \cmidrule(lr){5-7} \cmidrule(lr){8-10} \cmidrule(lr){11-13} \cmidrule(lr){14-15} 
   & & BLEU & ROUGE & P & R & F1 & P & R & F1 & P & R & F1 & BLEU & ROUGE \\
\midrule
% \multicolumn{2}{c}{} & BLEU & \multicolumn{1}{c}{ROUGE} & P & R & \multicolumn{1}{c}{F1} & P & R & \multicolumn{1}{c}{F1} & P & R & \multicolumn{1}{c}{F1} & BLEU & ROUGE \\ \hline
\multirow{3}{*}{Clarify} & yes-no & 65.3 & 67.9 & \textbf{46.9} & 15.0 & 22.7 & 22.4 & 20.0 & 10.9 & --- & --- & --- & \textbf{28.0} & \textbf{30.7} \\ 
 & choice & \textbf{74.3} & \textbf{75.9} & 22.4 & \textbf{37.4} & 28.0 & \textbf{46.4} & \textbf{38.0} & \textbf{29.6} & --- & --- & --- & 20.0 & 21.7 \\ 
 & open & 67.2 & 68.9 & 8.4 & 17.6 & 11.4 & 38.9 & 22.1 & 16.7 & --- & --- & --- & 15.0 & 16.3 \\ \midrule
\multirow{3}{*}{Answer} & opinion & 57.1 & 57.2 & 2.3 & 17.3 & 3.2 & --- & --- & --- & 32.3 & \textbf{37.0} & 23.3 & 12.6 & 13.6 \\ 
 & fact & 60.1 & 61.2 & 13.5 & 21.2 & 15.7 & --- & --- & --- & \textbf{44.1} & 34.6 & \textbf{30.0} & 16.4 & 19.6 \\ 
 & open & 62.7 & 62.8 & 39.1 & 35.6 & \textbf{36.7} & --- & --- & --- & 41.6 & 35.1 & 29.1 & 13.4 & 17.2 \\ 
\bottomrule
\end{tabular}
\end{table*}

From Table~\ref{tab: result}, there are two main observations.
First, WISE+GT achieves the best performance in terms of all key metrics. 
As we can see in Table~\ref{tab: result}, WISE+GT achieves 20.4\% and 24.2\% in terms of BLEU and ROUGE for \acs{RG}, outperforming the strongest model WISE by 5.8\% and 7.3\%, respectively.
This shows that the human labels of \acs{ID}, \acs{KE}, \acs{AP}, \acs{QS}, \acs{PS} do  benefit \acs{RG}. 

Second, \ac{WISE} achieves a higher performance than WISE-pretrain in terms of most metrics, indicating that the pretrain phrase is of great benefit for the performance of \ac{WISE}.
For example, the BLEU and ROUGE score of \ac{WISE} for \acs{RG} are 13.6\% and 15.9\%, much higher than those of WISE-pretrain, respectively.
The reason is that the model can leverage external knowledge in the pretrain phrase, which boosts the final performance.

From Table~\ref{tab: DiffAction}, we obtain three main findings.
First, for the action `Clarify', the best performance in terms of \acs{KE}, \acs{AP} and \acs{QS} is achieved w.r.t.\ `Clarify choice' and in terms of \acs{RG} w.r.t.\ `Clarify yes-no'.
The worst performance in terms of \acs{KE} and \acs{QS} is achieved w.r.t.\ `Clarify yes-no' and in terms of \acs{AP} and \acs{RG} w.r.t.\ `Clarify open'.
% For example, as shown in Table~\ref{tab: DiffAction}, WISE achieves the best performance~(28.0\% in terms of BLEU) for \acs{RG} w.r.t. `Clarify yes-no', but the worst performance~(10.9\% in terms of F1) for \acs{QS}.
For response generation w.r.t.\ `Clarify yes-no', WISE can almost copy the answer from a single query, which is easier than response generation w.r.t.\ `Clarify choice' and `Clarify open', which different queries, like `city' from `Beijing' and `Jinan'.

Second, for the action `Answer', the best performance in terms of \acs{KE} and \acs{AP} is achieved w.r.t.\ `Answer-open' and in terms of \acs{PS} and \acs{RG} w.r.t.\ `Answer-fact'.
The worst performance in terms of all tasks is achieved w.r.t.\ `Answer-opinion'.
% As seen in Table~\ref{tab: DiffAction}, the best performance~(16.4\% in terms of BLEU) in terms of \acs{RG} is achieved w.r.t. `Answer fact'.
For response generation w.r.t.\ `Answer fact', WISE can copy rather than summarize from multiple passages w.r.t.\ `Answer open' and `Answer opinion'. 

Third, the performance of \ac{PS} has a high correlation with the final \ac{RG} task w.r.t.\ `Answer', while the performance of \ac{QS} does not seem to be correlated w.r.t.\ `Clarify'.
As we can see in Table~\ref{tab: DiffAction}, an F1 score of 23.3\% in \ac{PS} and BLEU score of 12.6\% in terms of \ac{RG} is achieved w.r.t.\ `Answer opinion', and a higher F1 score~(30.0\%) in terms of \ac{PS} and a higher BLEU score~(16.4\%) in terms of \ac{RG} w.r.t.\ `Answer fact'.
For response generation w.r.t.\ `Answer', large portions of the output can be copied from passages, and thus the performance \ac{PS} is crucial to the final \ac{RG} performance.
The F1 score of WISE is 10.9\% in \ac{QS} w.r.t.\ `Clarify yes-no', which is lower than 29.6\% w.r.t.\ `Clarify choice'; the performance in \ac{RG} is better, e.g., 28\% vs.\ 20\%.
It is more likely that WISE has to copy from multiple queries for `Clarify choice' and from a single query for `Clarify yes-no', leading to performance differences for \ac{QS} and \ac{RG}, w.r.t.\ `Clarify'.

We split the test dataset based on the occurrence of conversational background in the training dataset: (test~(unseen) means no occurrence, and test~(seen) means occurrence); the results are shown in Table~\ref{tab: result_seen_unseen}.
Generally, the performances of WISE on all tasks is best on the test~(seen) dataset, and worst on the test~(unseen) dataset.
For example, it achieves 29.6\% in terms of F1 for \ac{PS} in test~(seen) dataset, which is higher than 24.9\% in test dataset and 20.5\% in test~(unseen) dataset.
WISE has been trained on the same conversational background in the test~(seen) condition, and has never seen the conversational background in the test~(unseen) condition.

\subsection{Effects of subtasks}
We analyze the effects of different tasks in WISE, w.r.t.\ \acs{ID}, \acs{KE}, \acs{AP}, \acs{QS} and \acs{PS}, and the results are shown in Table~\ref{tab: DiffTask}. 
We remove each task, resulting in five settings; `-xx' means \ac{WISE} without module `xx'. 
% first view
Generally, removing any module will result in a drop in performance for \ac{RG} on all metrics, indicating that all parts are helpful to WISE. 
% analyze data
Without \ac{QS}, \ac{RG} drops sharply on all metrics.
Specially, it drops 4.2\% in terms of BLEU for \ac{RG}, which means \ac{QS} is essential to \acs{RG}. 
While without \ac{PS}, \ac{RG} drops a little compared to \ac{QS}.
Specially, it drops only 0.9\% in terms of BLEU for \ac{RG}.
The performance of \ac{PS} is far from perfect, only 24.9\% in F1 score; omitting it has limited impact.
In addition, removing \acs{ID}, \acs{KE} or \acs{AP} harms the performance of \ac{QS} and \ac{PS}, indicating that all previous modules for \ac{QS} and \ac{PS} help for the performance of \ac{QS} and \ac{PS}.
% analyze data
Specifically, performance drops by 10\% and 5.7\% in terms of F1 score for \ac{QS} and \ac{PS} when removing \ac{KE}: if the model does not extract the accurate keyphrase, it can not select the most relevant queries and passages.

\begin{table*}[ht]
\centering
\caption{Overall performance of models removing different tasks.}
\label{tab: DiffTask} 

\begin{tabular}{lcccccccccccccccc}
\toprule
     & \multicolumn{3}{c}{ID~(\%)} & \multicolumn{2}{c}{KE~(\%)} & \multicolumn{3}{c}{AP~(\%)} & \multicolumn{3}{c}{QS~(\%)} & \multicolumn{3}{c}{PS~(\%)} & \multicolumn{2}{c}{RG~(\%)} \\
\cmidrule(lr){2-4} \cmidrule(lr){5-6} \cmidrule(lr){7-9} \cmidrule(lr){10-12} \cmidrule(lr){13-15} \cmidrule(lr){16-17}
     & P     & R     & F1     & BLEU      & ROUGE      & P     & R     & F1     & P     & R     & F1     & P     & R     & F1     & BLEU      & ROUGE      \\
\midrule
-\acs{ID} &  ---    &   ---    &    ---    &     54.6      &      54.8      &   18.7    &   22.6    &    18.3    &   25.5    &   8.3    &  6.1      &    \textbf{43.5}   &   25.1    &   22.7     &    13.0       &    15.2        \\

-\acs{KE}     &   \textbf{66.2}    &   28.2    &   30.6     &    ---       &      ---      &   22.0    &   22.7    &   \textbf{19.1}     &   \textbf{28.0}    &   1.6    &    1.5    &    39.8   &   20.8    &    19.2    &    12.6       &  14.7  \\  

-\acs{AP}     &   52.5    &  32.2     &    \textbf{35.3}    &    \textbf{59.5}       &       \textbf{59.4}     &   ---    &    ---   &   ---    &   21.6    &   11.0    &    8.5    &   37.8    &    26.4   &    21.9    &    13.2       &  15.1  \\

-\acs{QS}     &   51.9    &   \textbf{32.6}    &    32.6    &    54.2       &       54.2     &   \textbf{22.7}    &    \textbf{23.2}   &   18.9     &   ---   &   ---    &   ---    &  39.5     &   25.9    &    22.1    &     10.4      &  12.1  \\

-\acs{PS}     &  51.3     &    30.8   &  32.8      &    56.8       &   57.0         &  20.1     &  22.3     &    18.1    &   24.7    &    8.1   &     6.3   &    ---   &   ---    &    ---    &    13.7  &  15.7  \\

WISE    &    45.2   &    32.5   &  34.1      &    58.3       & 59.0    &   18.8    &  20.6      &    17.8   &  26.9     &  \textbf{16.1}      &    \textbf{11.5}   &   35.8    &  \textbf{31.5}      &    \textbf{24.9}       &  \textbf{14.6} & \textbf{16.9}  \\ 
\bottomrule
\end{tabular}

\end{table*}

\begin{table*}[th]
\centering
\caption{Overall performance of \acs{WISE} removing different datasets for pretraining.}
\label{tab: DiffDataset} 

\begin{tabular}{lcccccccccccccccc}
\toprule
     & \multicolumn{3}{c}{ID~(\%)} & \multicolumn{2}{c}{KE~(\%)} & \multicolumn{3}{c}{AP~(\%)} & \multicolumn{3}{c}{QS~(\%)} & \multicolumn{3}{c}{PS~(\%)} & \multicolumn{2}{c}{RG~(\%)} \\
\cmidrule(lr){2-4} \cmidrule(lr){5-6} \cmidrule(lr){7-9} \cmidrule(lr){10-12} \cmidrule(lr){13-15} \cmidrule(lr){16-17}
     & P     & R     & F1     & BLEU      & ROUGE      & P     & R     & F1     & P     & R     & F1     & P     & R     & F1     & BLEU      & ROUGE      \\
\midrule
-DuReader &   \textbf{47.5}    &  25.7     &  27.7      &      50.7     &        51.8    &   18.0    &    19.5   &   17.2     &   \textbf{33.3}    &  2.2     &   2.4    &   33.7    &    17.5    &     15.1      &      9.6 & 11.0      \\

-KdConv &   41.1    &   27.7    &   28.1     &    49.7       &       51.3     &   16.3    &  17.7     &   15.3     &   22.2    &    8.4   &    6.3    &    \textbf{37.2}   &   26.0    &   22.8     &     12.5      &    14.7        \\

-DuConv     &   43.9    &    \textbf{35.5}   &    \textbf{35.8}    &    55.8       &     57.0       &    20.3   &   20.2    &   17.9     &   30.8    &     7.7  &   6.6     &  34.8     &   28.0    &  22.6      &    13.4       &   15.4 \\  

-WebQA     &   39.0    &    30.6   &   32.0     &    57.1       &     57.4       &   \textbf{20.9}    &    \textbf{20.9}   &     \textbf{18.8}   &   27.7    &   10.6    &    8.4    &    34.9   &   23.6    &   19.6     &   12.3        &  14.3  \\

WISE     &    45.2   &    32.5   &  34.1      &    \textbf{58.3}       & \textbf{59.0}    &   18.8    &  20.6      &    17.8   &  26.9     &  \textbf{16.1}      &    \textbf{11.5}   &   35.8    &  \textbf{31.5}      &    \textbf{24.9}       &  \textbf{14.6} & \textbf{16.9}  \\ 
\bottomrule
\end{tabular}

\end{table*}

\subsection{Effects of pretraining}
To analyze the effects of different pretrain datasets, we conduct an ablation study~(Table~\ref{tab: DiffDataset}).
WISE uses four external datasets for pretraining; each time we remove one,  we use the others for pretraining. Hence, we have four conditions.
In Table~\ref{tab: DiffDataset}, `-xx' means pretraining \ac{WISE} using datasets other than `xx'.

The results show that all datasets are helpful for pretraining; removing any dataset results in a performance drop for \ac{RG}. 
The results drop in terms of all metrics when pretraining without the DuReader dataset. 
%As we can see from Table~\ref{tab: DiffDataset}, 
The results drop by 5.0\% and 5.9\% in terms of BLEU and ROUGE for \ac{RG}.
The drop after removing pretraining with DuReader is far larger than after removing any of the other individual pretraining dataset, indicating the importance of DuReader for pretraining, as shown in Table~\ref{table_pre-training}.
Interestingly, sometimes removing a dataset will improve the performance of \ac{ID} and \ac{AP}.
For example, -duconv achieves 35.8\% in terms of F1 in \ac{ID}, higher than the 34.1\% for \ac{WISE}.
\ac{ID} and \ac{AP} predict labels among small sets, which are relatively easy compared to the other tasks.
%And it influences their performance when the model pretrains other tasks with corresponding datasets.

% !TEX root =  ../main.tex

\section{Related Work}
% We survey related works from three perspectives: user studies, datasets, and frameworks.

\paragraph{User studies}
To determine whether \ac{CIS} is needed and what it should look like, researchers have conducted different user studies.
\citet{Vtyurina:2017:ECS:3027063.3053175} ask 21 participants to solve 3 information seeking tasks by conversing with three agents: an existing commercial system, a human expert, and a perceived experimental automatic system, backed by a human ``wizard behind the curtain.''
They conclude that people do not have biases against \ac{CIS} systems, as long as their performance is acceptable.
% , and existing conversational assistants are not yet up to task, i.e., they cannot be effectively used for complex information search tasks.
%
% \citet{Trippas:2018:IDS:3176349.3176387} conduct a laboratory-based observational study of 13 pairs of participants each completing three search tasks.
% They identify three themes in \ac{CIS}: meta-communication (Intermediaries engage in communication about the information request with seekers), SERP (Intermediaries present the SERP to seekers), and scanning document (Intermediaries access a document and present the content to seekers).
%
\citet{Trippas:2018:IDS:3176349.3176387} conduct a laboratory-based observational study, where pairs of people perform search tasks communicating verbally.
They conclude that \ac{CIS} is more complex and interactive than traditional search.
Based on their conclusions, we can confirm that \ac{CIS} is the right direction but there is still a long way to go.

\paragraph{Datasets}
Several conversation related datasets are available, for various tasks, e.g., knowledge grounded conversations~\cite{moghe2018towards,dinan2018wizard}, conversational question answering~\cite{10.1145/3357384.3358016,DBLP:conf/aaai/SahaPKSC18,choi-etal-2018-quac,reddy-etal-2019-coqa}.
However, their targets are different from information seeking in the search scenarios.

To address this gap, there are various efforts to build datasets for \ac{CIS}.
\citet{cast2019} release the CAsT dataset, which aims to establish a concrete and standard collection of data to make \ac{CIS} systems directly comparable.
% Currently they only target two tasks: 
% (1) Read the dialogue context: Track the evolution of the information need in the conversation, identifying salient information needed for the current turn in the conversation, similar to \ac{KE} in this work.
% (2) Retrieve candidate response information: Perform retrieval over a large collection of paragraphs, similar to \ac{PS} in this work.
\citet{ren-2020-conversations-arxiv} provide natural language responses by summarizing relevant information in paragraphs based on CAsT.
\citet{Aliannejadi:2019:ACQ:3331184.3331265} explore how to ask clarifying questions, which is important in order to achieve mixed-initiative \ac{CIS}.
They propose an offline evaluation methodology for the task and collect the dataset ClariQ through crowdsourcing.
The above datasets only cover some aspects of \ac{CIS}.

To better understand \ac{CIS} conversations, \citet{10.1145/3176349.3176388} release the MISC dataset, which is built by recording pairs of a seeker and intermediary collaborating on \ac{CIS}.
% They study the style of a conversation (pleasant, abrupt, confusing, courteous), and indications of style which can be computed at scale.
The work most closely related to ours is by \citet{TRIPPAS2020102162}.
They create the CSCData dataset for spoken conversational search (very close to the \ac{CIS} setting), define a labelling set identifying the intents and actions of seekers and intermediaries, and conduct a deep analysis.
The MISC and CSCData datasets only have 666 and 1,044 utterances, which results in a low coverage of search intents and answer types.
No supporting documents are available and the responses are in very verbose spoken language.
Importantly, it is hard for the defined labels to make a direct connection to \acp{TSE}.

\paragraph{Frameworks}
Several recent frameworks address a specific aspect of \ac{CIS}, e.g., asking clarifying questions~\cite{Hamedwww2020}, generating responses for natural language questions~\cite{conf/nips/NguyenRSGTMD16}.
There are also attempts to design a complete \ac{CIS} that could work in practice.
\citet{10.1145/3020165.3020183} consider the question of what properties would be desirable for a \ac{CIS} system so that the system enables users to answer a variety of information need in a natural and efficient manner.
They present a theoretical framework of information interaction in a chat setting for \ac{CIS}, which gives guidelines for designing a practical \ac{CIS} system.
\citet{azzopardi2018conceptualizing} outline the actions and intents of users and systems explaining how these actions enable users to explore the search space and resolve their information need.
Their work provides a conceptualization of the \ac{CIS} process and a framework for the development of practical \ac{CIS} systems.

To sum up, the work listed above either focuses on a particular aspect or studies the modeling of \ac{CIS} theoretically.
% Currently, there is no workable model that is able to fully replace the intermediary.
% The proposed frameworks have made requirements about the data annotations more demanding, often going beyond currently available datasets.

% !TEX root =  ../main.tex

\section{Conclusion and future work}
We have proposed a pipeline and a modular end-to-end neural architecture for \acf{CIS}, which models \ac{CIS} as six sub-tasks to improve the performance on the response generation (RG) task.
We also collected a \ac{WISE} dataset in a wizard-of-oz fashion, which is more suitable and challenging for comprehensive and in-depth research on all aspects of \ac{CIS}.
Extensive experiments on the \ac{WISE} dataset show that the proposed \acs{WISE} model can achieve state-of-the-art performance and each proposed module of \ac{WISE} contributes to the final response generation.

Although we have proposed a standard dataset for \ac{CIS}, we have a long way to go to improve the performance on each subtask, and to consider more factors, e.g., the information timeliness~\cite{10.1145/3309547}. 
A potential direction is to leverage large-scale unlabeled datasets on related tasks like dialogue and information retrieval.

\section*{Reproducibility}
The code and data used to produce the results in this paper are available at \url{https://github.com/PengjieRen/CaSE_WISE}.

\appendix

% !TEX root =  ../main.tex

% \begin{figure*}[ht]
% \centering
% \includegraphics[width=0.9\textwidth]{figures/UI-1.png}
% \caption{Interface for the searcher.}
% \label{figure_UI1}
% \end{figure*}

% \begin{figure*}[ht]
% \centering
% \includegraphics[width=0.9\textwidth]{figures/UI-2.png}
% \caption{Interface for the wizard.}
% \label{figure_UI2}
% \end{figure*}
 
\begin{acks}
We thank the reviewers for their valuable feedback.
This research was partially supported by
the National Key R\&D Program of China with grant No. 2020YFB1406704,
the Natural Science Foundation of China (61972234, 61902219, 62072279), 
the Key Scientific and Technological Innovation Program of Shandong Province (2019JZZY010129), 
the Tencent WeChat Rhino-Bird Focused Research Program (JR-WXG-2021411),
the Fundamental Research Funds of Shandong University,
and 
the Hybrid Intelligence Center, a 10-year programme funded by the Dutch Ministry of Education, Culture and Science through the Netherlands Organisation for Scientific Research, \url{https://hybrid-intelligence-centre.nl}.
All content represents the opinion of the authors, which is not necessarily shared or endorsed by their respective employers and/or sponsors.
\end{acks}

\bibliographystyle{ACM-Reference-Format}
\bibliography{references}


\begin{thebibliography}{49}

%%% the Web address.

\ifx \showCODEN    \undefined \def \showCODEN     #1{\unskip}     \fi
\ifx \showDOI      \undefined \def \showDOI       #1{#1}\fi
\ifx \showISBNx    \undefined \def \showISBNx     #1{\unskip}     \fi
\ifx \showISBNxiii \undefined \def \showISBNxiii  #1{\unskip}     \fi
\ifx \showISSN     \undefined \def \showISSN      #1{\unskip}     \fi
\ifx \showLCCN     \undefined \def \showLCCN      #1{\unskip}     \fi
\ifx \shownote     \undefined \def \shownote      #1{#1}          \fi
\ifx \showarticletitle \undefined \def \showarticletitle #1{#1}   \fi
\ifx \showURL      \undefined \def \showURL       {\relax}        \fi
% invisible to TeX
\providecommand\bibfield[2]{#2}
\providecommand\bibinfo[2]{#2}
\providecommand\natexlab[1]{#1}
\providecommand\showeprint[2][]{arXiv:#2}

\bibitem[\protect\citeauthoryear{Agarap}{Agarap}{2018}]%
        {DBLP:journals/corr/abs-1803-08375}
\bibfield{author}{\bibinfo{person}{Abien~Fred Agarap}.}
  \bibinfo{year}{2018}\natexlab{}.
\newblock \showarticletitle{Deep Learning using Rectified Linear Units (ReLU)}.
\newblock \bibinfo{journal}{\emph{arXiv preprint arXiv:1803.08375}}
  (\bibinfo{year}{2018}).
\newblock


\bibitem[\protect\citeauthoryear{Aliannejadi, Zamani, Crestani, and
  Croft}{Aliannejadi et~al\mbox{.}}{2019}]%
        {Aliannejadi:2019:ACQ:3331184.3331265}
\bibfield{author}{\bibinfo{person}{Mohammad Aliannejadi},
  \bibinfo{person}{Hamed Zamani}, \bibinfo{person}{Fabio Crestani}, {and}
  \bibinfo{person}{W.~Bruce Croft}.} \bibinfo{year}{2019}\natexlab{}.
\newblock \showarticletitle{Asking Clarifying Questions in Open-Domain
  Information-Seeking Conversations}. In \bibinfo{booktitle}{\emph{Proceedings
  of the 42nd International ACM SIGIR Conference on Research and Development in
  Information Retrieval, {SIGIR} 2019}}. \bibinfo{pages}{475--484}.
\newblock


\bibitem[\protect\citeauthoryear{Anand, Cavedon, Hagen, Joho, Sanderson, and
  Stein}{Anand et~al\mbox{.}}{2020}]%
        {DBLP:journals/corr/abs-2005-08658}
\bibfield{author}{\bibinfo{person}{Avishek Anand}, \bibinfo{person}{Lawrence
  Cavedon}, \bibinfo{person}{Matthias Hagen}, \bibinfo{person}{Hideo Joho},
  \bibinfo{person}{Mark Sanderson}, {and} \bibinfo{person}{Benno Stein}.}
  \bibinfo{year}{2020}\natexlab{}.
\newblock \showarticletitle{Conversational Search - {A} Report from Dagstuhl
  Seminar 19461}.
\newblock \bibinfo{journal}{\emph{arXiv preprint arXiv:2005.08658}}
  (\bibinfo{year}{2020}).
\newblock


\bibitem[\protect\citeauthoryear{Azzopardi, Dubiel, Halvey, and
  Dalton}{Azzopardi et~al\mbox{.}}{2018}]%
        {azzopardi2018conceptualizing}
\bibfield{author}{\bibinfo{person}{Leif Azzopardi}, \bibinfo{person}{Mateusz
  Dubiel}, \bibinfo{person}{Martin Halvey}, {and} \bibinfo{person}{Jeffery
  Dalton}.} \bibinfo{year}{2018}\natexlab{}.
\newblock \showarticletitle{Conceptualizing Agent-human Interactions during the
  Conversational Search Process}. In \bibinfo{booktitle}{\emph{The 2nd Workshop
  on Conversational Approaches to Information Retrieval}}.
\newblock


\bibitem[\protect\citeauthoryear{Ba, Kiros, and Hinton}{Ba
  et~al\mbox{.}}{2016}]%
        {ba2016layer}
\bibfield{author}{\bibinfo{person}{Lei~Jimmy Ba}, \bibinfo{person}{Jamie~Ryan
  Kiros}, {and} \bibinfo{person}{Geoffrey~E. Hinton}.}
  \bibinfo{year}{2016}\natexlab{}.
\newblock \showarticletitle{Layer Normalization}.
\newblock \bibinfo{journal}{\emph{arXiv preprint arXiv:1607.06450}}
  (\bibinfo{year}{2016}).
\newblock


\bibitem[\protect\citeauthoryear{Baheti, Ritter, and Small}{Baheti
  et~al\mbox{.}}{2020}]%
        {baheti-etal-2020-fluent}
\bibfield{author}{\bibinfo{person}{Ashutosh Baheti}, \bibinfo{person}{Alan
  Ritter}, {and} \bibinfo{person}{Kevin Small}.}
  \bibinfo{year}{2020}\natexlab{}.
\newblock \showarticletitle{Fluent Response Generation for Conversational
  Question Answering}. In \bibinfo{booktitle}{\emph{Proceedings of the 58th
  Annual Meeting of the Association for Computational Linguistics, {ACL}
  2020}}. \bibinfo{pages}{191--207}.
\newblock


\bibitem[\protect\citeauthoryear{Broder}{Broder}{2002}]%
        {10.1145/792550.792552}
\bibfield{author}{\bibinfo{person}{Andrei~Z. Broder}.}
  \bibinfo{year}{2002}\natexlab{}.
\newblock \showarticletitle{A taxonomy of web search}.
\newblock \bibinfo{journal}{\emph{{SIGIR} Forum}}  \bibinfo{volume}{36}
  (\bibinfo{year}{2002}), \bibinfo{pages}{3--10}.
\newblock


\bibitem[\protect\citeauthoryear{Choi, He, Iyyer, Yatskar, Yih, Choi, Liang,
  and Zettlemoyer}{Choi et~al\mbox{.}}{2018}]%
        {choi-etal-2018-quac}
\bibfield{author}{\bibinfo{person}{Eunsol Choi}, \bibinfo{person}{He He},
  \bibinfo{person}{Mohit Iyyer}, \bibinfo{person}{Mark Yatskar},
  \bibinfo{person}{Wen{-}tau Yih}, \bibinfo{person}{Yejin Choi},
  \bibinfo{person}{Percy Liang}, {and} \bibinfo{person}{Luke Zettlemoyer}.}
  \bibinfo{year}{2018}\natexlab{}.
\newblock \showarticletitle{QuAC: Question Answering in Context}. In
  \bibinfo{booktitle}{\emph{Proceedings of the 2018 Conference on Empirical
  Methods in Natural Language Processing, {EMNLP} 2018}}.
  \bibinfo{pages}{2174--2184}.
\newblock


\bibitem[\protect\citeauthoryear{Christakopoulou, Radlinski, and
  Hofmann}{Christakopoulou et~al\mbox{.}}{2016}]%
        {10.1145/2939672.2939746}
\bibfield{author}{\bibinfo{person}{Konstantina Christakopoulou},
  \bibinfo{person}{Filip Radlinski}, {and} \bibinfo{person}{Katja Hofmann}.}
  \bibinfo{year}{2016}\natexlab{}.
\newblock \showarticletitle{Towards Conversational Recommender Systems}. In
  \bibinfo{booktitle}{\emph{Proceedings of the 22nd ACM SIGKDD International
  Conference on Knowledge Discovery and Data Mining, {SIGKDD} 2016}}.
  \bibinfo{pages}{815--824}.
\newblock


\bibitem[\protect\citeauthoryear{Christmann, Roy, Abujabal, Singh, and
  Weikum}{Christmann et~al\mbox{.}}{2019}]%
        {10.1145/3357384.3358016}
\bibfield{author}{\bibinfo{person}{Philipp Christmann},
  \bibinfo{person}{Rishiraj~Saha Roy}, \bibinfo{person}{Abdalghani Abujabal},
  \bibinfo{person}{Jyotsna Singh}, {and} \bibinfo{person}{Gerhard Weikum}.}
  \bibinfo{year}{2019}\natexlab{}.
\newblock \showarticletitle{Look before You Hop: Conversational Question
  Answering over Knowledge Graphs Using Judicious Context Expansion}. In
  \bibinfo{booktitle}{\emph{Proceedings of the 28th ACM International
  Conference on Information and Knowledge Management, {CIKM} 2019}}.
  \bibinfo{pages}{729--738}.
\newblock


\bibitem[\protect\citeauthoryear{Czyzewski, Dalton, and Leuski}{Czyzewski
  et~al\mbox{.}}{2020}]%
        {10.1145/3397271.3401397}
\bibfield{author}{\bibinfo{person}{Adam Czyzewski}, \bibinfo{person}{Jeffrey
  Dalton}, {and} \bibinfo{person}{Anton Leuski}.}
  \bibinfo{year}{2020}\natexlab{}.
\newblock \showarticletitle{Agent Dialogue: {A} Platform for Conversational
  Information Seeking Experimentation}. In
  \bibinfo{booktitle}{\emph{Proceedings of the 43rd International ACM SIGIR
  conference on research and development in Information Retrieval, {SIGIR}
  2020}}. \bibinfo{pages}{2121--2124}.
\newblock


\bibitem[\protect\citeauthoryear{Dalton, Xiong, and Callan}{Dalton
  et~al\mbox{.}}{2020a}]%
        {cast2019}
\bibfield{author}{\bibinfo{person}{Jeffrey Dalton}, \bibinfo{person}{Chenyan
  Xiong}, {and} \bibinfo{person}{Jamie Callan}.}
  \bibinfo{year}{2020}\natexlab{a}.
\newblock \showarticletitle{{TREC} CAsT 2019: The Conversational Assistance
  Track Overview}.
\newblock \bibinfo{journal}{\emph{arXiv preprint arXiv:2003.13624}}
  (\bibinfo{year}{2020}).
\newblock


\bibitem[\protect\citeauthoryear{Dalton, Xiong, Kumar, and Callan}{Dalton
  et~al\mbox{.}}{2020b}]%
        {10.1145/3397271.3401206}
\bibfield{author}{\bibinfo{person}{Jeffrey Dalton}, \bibinfo{person}{Chenyan
  Xiong}, \bibinfo{person}{Vaibhav Kumar}, {and} \bibinfo{person}{Jamie
  Callan}.} \bibinfo{year}{2020}\natexlab{b}.
\newblock \showarticletitle{CAsT-19: {A} Dataset for Conversational Information
  Seeking}. In \bibinfo{booktitle}{\emph{Proceedings of the 43rd International
  ACM SIGIR conference on research and development in Information Retrieval,
  {SIGIR} 2020}}. \bibinfo{pages}{1985--1988}.
\newblock


\bibitem[\protect\citeauthoryear{Dinan, Roller, Shuster, Fan, Auli, and
  Weston}{Dinan et~al\mbox{.}}{2019}]%
        {dinan2018wizard}
\bibfield{author}{\bibinfo{person}{Emily Dinan}, \bibinfo{person}{Stephen
  Roller}, \bibinfo{person}{Kurt Shuster}, \bibinfo{person}{Angela Fan},
  \bibinfo{person}{Michael Auli}, {and} \bibinfo{person}{Jason Weston}.}
  \bibinfo{year}{2019}\natexlab{}.
\newblock \showarticletitle{Wizard of Wikipedia: Knowledge-Powered
  Conversational Agents}. In \bibinfo{booktitle}{\emph{Proceedings of the 7th
  International Conference on Learning Representations, {ICLR} 2019}}.
\newblock


\bibitem[\protect\citeauthoryear{Feng, He, Wang, Luo, Liu, and Chua}{Feng
  et~al\mbox{.}}{2019}]%
        {10.1145/3309547}
\bibfield{author}{\bibinfo{person}{Fuli Feng}, \bibinfo{person}{Xiangnan He},
  \bibinfo{person}{Xiang Wang}, \bibinfo{person}{Cheng Luo},
  \bibinfo{person}{Yiqun Liu}, {and} \bibinfo{person}{Tat-Seng Chua}.}
  \bibinfo{year}{2019}\natexlab{}.
\newblock \showarticletitle{Temporal Relational Ranking for Stock Prediction}.
\newblock \bibinfo{journal}{\emph{ACM Trans. Inf. Syst.}} \bibinfo{volume}{37},
  \bibinfo{number}{2} (\bibinfo{year}{2019}), \bibinfo{numpages}{30}~pages.
\newblock


\bibitem[\protect\citeauthoryear{Gao, Lei, He, de~Rijke, and Chua}{Gao
  et~al\mbox{.}}{2021}]%
        {gao-2021-advances-arxiv}
\bibfield{author}{\bibinfo{person}{Chongming Gao}, \bibinfo{person}{Wenqiang
  Lei}, \bibinfo{person}{Xiangnan He}, \bibinfo{person}{Maarten de Rijke},
  {and} \bibinfo{person}{Tat-Seng Chua}.} \bibinfo{year}{2021}\natexlab{}.
\newblock \showarticletitle{Advances and Challenges in Conversational
  Recommender Systems: A Survey}.
\newblock \bibinfo{journal}{\emph{arXiv preprint arXiv:2101.09459}}
  (\bibinfo{year}{2021}).
\newblock


\bibitem[\protect\citeauthoryear{He, Liu, Liu, Lyu, Zhao, Xiao, Liu, Wang, Wu,
  She, Liu, Wu, and Wang}{He et~al\mbox{.}}{2018}]%
        {he-etal-2018-dureader}
\bibfield{author}{\bibinfo{person}{Wei He}, \bibinfo{person}{Kai Liu},
  \bibinfo{person}{Jing Liu}, \bibinfo{person}{Yajuan Lyu},
  \bibinfo{person}{Shiqi Zhao}, \bibinfo{person}{Xinyan Xiao},
  \bibinfo{person}{Yuan Liu}, \bibinfo{person}{Yizhong Wang},
  \bibinfo{person}{Hua Wu}, \bibinfo{person}{Qiaoqiao She},
  \bibinfo{person}{Xuan Liu}, \bibinfo{person}{Tian Wu}, {and}
  \bibinfo{person}{Haifeng Wang}.} \bibinfo{year}{2018}\natexlab{}.
\newblock \showarticletitle{DuReader: a Chinese Machine Reading Comprehension
  Dataset from Real-world Applications}. In
  \bibinfo{booktitle}{\emph{Proceedings of the Workshop on Machine Reading for
  Question Answering, {ACL} 2018}}. \bibinfo{pages}{37--46}.
\newblock


\bibitem[\protect\citeauthoryear{Kocisk{\'{y}}, Schwarz, Blunsom, Dyer,
  Hermann, Melis, and Grefenstette}{Kocisk{\'{y}} et~al\mbox{.}}{2018}]%
        {journals/tacl/KociskySBDHMG18}
\bibfield{author}{\bibinfo{person}{Tom{\'{a}}s Kocisk{\'{y}}},
  \bibinfo{person}{Jonathan Schwarz}, \bibinfo{person}{Phil Blunsom},
  \bibinfo{person}{Chris Dyer}, \bibinfo{person}{Karl~Moritz Hermann},
  \bibinfo{person}{G{\'{a}}bor Melis}, {and} \bibinfo{person}{Edward
  Grefenstette}.} \bibinfo{year}{2018}\natexlab{}.
\newblock \showarticletitle{The NarrativeQA Reading Comprehension Challenge}.
\newblock \bibinfo{journal}{\emph{Trans. Assoc. Comput. Linguistics}}
  \bibinfo{volume}{6} (\bibinfo{year}{2018}), \bibinfo{pages}{317--328}.
\newblock


\bibitem[\protect\citeauthoryear{Kratzwald and Feuerriegel}{Kratzwald and
  Feuerriegel}{2019}]%
        {10.1145/3308558.3313661}
\bibfield{author}{\bibinfo{person}{Bernhard Kratzwald} {and}
  \bibinfo{person}{Stefan Feuerriegel}.} \bibinfo{year}{2019}\natexlab{}.
\newblock \showarticletitle{Learning from On-Line User Feedback in Neural
  Question Answering on the Web}. In \bibinfo{booktitle}{\emph{Proceedings of
  the World Wide Web Conference, {WWW} 2019}}. \bibinfo{pages}{906--916}.
\newblock


\bibitem[\protect\citeauthoryear{Li, Li, He, Wang, Cao, Zhou, and Xu}{Li
  et~al\mbox{.}}{2016}]%
        {DBLP:journals/corr/LiLHWCZX16}
\bibfield{author}{\bibinfo{person}{Peng Li}, \bibinfo{person}{W. Li},
  \bibinfo{person}{Z. He}, \bibinfo{person}{Xuguang Wang}, \bibinfo{person}{Y.
  Cao}, \bibinfo{person}{J. Zhou}, {and} \bibinfo{person}{W. Xu}.}
  \bibinfo{year}{2016}\natexlab{}.
\newblock \showarticletitle{Dataset and Neural Recurrent Sequence Labeling
  Model for Open-Domain Factoid Question Answering}.
\newblock \bibinfo{journal}{\emph{arXiv preprint arXiv:1607.06275}}
  (\bibinfo{year}{2016}).
\newblock


\bibitem[\protect\citeauthoryear{Lin}{Lin}{2004}]%
        {lin-2004-rouge}
\bibfield{author}{\bibinfo{person}{Chin-Yew Lin}.}
  \bibinfo{year}{2004}\natexlab{}.
\newblock \showarticletitle{Rouge: A Package for Automatic Evaluation of
  Summaries}. In \bibinfo{booktitle}{\emph{Proceedings of the 42nd Annual
  Meeting of the Association for Computational Linguistics, {ACL} 2002}}.
  \bibinfo{pages}{74--81}.
\newblock


\bibitem[\protect\citeauthoryear{Moghe, Arora, Banerjee, and Khapra}{Moghe
  et~al\mbox{.}}{2018}]%
        {moghe2018towards}
\bibfield{author}{\bibinfo{person}{Nikita Moghe}, \bibinfo{person}{Siddhartha
  Arora}, \bibinfo{person}{Suman Banerjee}, {and} \bibinfo{person}{Mitesh~M.
  Khapra}.} \bibinfo{year}{2018}\natexlab{}.
\newblock \showarticletitle{Towards Exploiting Background Knowledge for
  Building Conversation Systems}. In \bibinfo{booktitle}{\emph{Proceedings of
  the 2018 Conference on Empirical Methods in Natural Language Processing,
  Brussels, {EMNLP} 2018}}. \bibinfo{pages}{2322--2332}.
\newblock


\bibitem[\protect\citeauthoryear{Nguyen, Rosenberg, Song, Gao, Tiwary,
  Majumder, and Deng}{Nguyen et~al\mbox{.}}{2016}]%
        {conf/nips/NguyenRSGTMD16}
\bibfield{author}{\bibinfo{person}{Tri Nguyen}, \bibinfo{person}{Mir
  Rosenberg}, \bibinfo{person}{Xia Song}, \bibinfo{person}{Jianfeng Gao},
  \bibinfo{person}{Saurabh Tiwary}, \bibinfo{person}{Rangan Majumder}, {and}
  \bibinfo{person}{Li Deng}.} \bibinfo{year}{2016}\natexlab{}.
\newblock \showarticletitle{{MS} {MARCO:} {A} Human Generated MAchine Reading
  COmprehension Dataset}. In \bibinfo{booktitle}{\emph{Proceedings of the
  Workshop on Cognitive Computation: Integrating neural and symbolic approaches
  2016 co-located with the 30th Annual Conference on Neural Information
  Processing Systems {NIPS} 2016}}, Vol.~\bibinfo{volume}{1773}.
\newblock


\bibitem[\protect\citeauthoryear{Papineni, Roukos, Ward, and Zhu}{Papineni
  et~al\mbox{.}}{2002}]%
        {conf/acl/PapineniRWZ02}
\bibfield{author}{\bibinfo{person}{Kishore Papineni}, \bibinfo{person}{Salim
  Roukos}, \bibinfo{person}{Todd Ward}, {and} \bibinfo{person}{Wei{-}Jing
  Zhu}.} \bibinfo{year}{2002}\natexlab{}.
\newblock \showarticletitle{Bleu: a Method for Automatic Evaluation of Machine
  Translation}. In \bibinfo{booktitle}{\emph{Proceedings of the 40th Annual
  Meeting of the Association for Computational Linguistics, {ACL} 2002}}.
  \bibinfo{pages}{311--318}.
\newblock


\bibitem[\protect\citeauthoryear{Pei, Ren, Monz, and de~Rijke}{Pei
  et~al\mbox{.}}{2020}]%
        {pei-2020-retrospective}
\bibfield{author}{\bibinfo{person}{Jiahuan Pei}, \bibinfo{person}{Pengjie Ren},
  \bibinfo{person}{Christof Monz}, {and} \bibinfo{person}{Maarten de Rijke}.}
  \bibinfo{year}{2020}\natexlab{}.
\newblock \showarticletitle{Retrospective and Prospective Mixture-of-Generators
  for Task-Oriented Dialogue Response Generation}. In
  \bibinfo{booktitle}{\emph{Proceedings of the 24th European Conference on
  Artificial Intelligence, {ECAI} 2020}}, Vol.~\bibinfo{volume}{325}.
  \bibinfo{pages}{2148--2155}.
\newblock


\bibitem[\protect\citeauthoryear{Peng, Li, Li, Gao, {\c{C}}elikyilmaz, Lee, and
  Wong}{Peng et~al\mbox{.}}{2017}]%
        {peng-etal-2017-composite}
\bibfield{author}{\bibinfo{person}{Baolin Peng}, \bibinfo{person}{Xiujun Li},
  \bibinfo{person}{Lihong Li}, \bibinfo{person}{Jianfeng Gao},
  \bibinfo{person}{Asli {\c{C}}elikyilmaz}, \bibinfo{person}{Sungjin Lee},
  {and} \bibinfo{person}{Kam{-}Fai Wong}.} \bibinfo{year}{2017}\natexlab{}.
\newblock \showarticletitle{Composite Task-Completion Dialogue Policy Learning
  via Hierarchical Deep Reinforcement Learning}. In
  \bibinfo{booktitle}{\emph{Proceedings of the 2017 Conference on Empirical
  Methods in Natural Language Processing, {EMNLP} 2017}}.
  \bibinfo{pages}{2231--2240}.
\newblock


\bibitem[\protect\citeauthoryear{Qu, Yang, Croft, Zhang, Trippas, and Qiu}{Qu
  et~al\mbox{.}}{2019}]%
        {10.1145/3295750.3298924}
\bibfield{author}{\bibinfo{person}{Chen Qu}, \bibinfo{person}{Liu Yang},
  \bibinfo{person}{W.~Bruce Croft}, \bibinfo{person}{Yongfeng Zhang},
  \bibinfo{person}{Johanne~R. Trippas}, {and} \bibinfo{person}{Minghui Qiu}.}
  \bibinfo{year}{2019}\natexlab{}.
\newblock \showarticletitle{User Intent Prediction in Information-seeking
  Conversations}. In \bibinfo{booktitle}{\emph{Proceedings of the 2019
  Conference on Human Information Interaction and Retrieval, {CHIIR} 2019}}.
  \bibinfo{pages}{25--33}.
\newblock


\bibitem[\protect\citeauthoryear{Radlinski and Craswell}{Radlinski and
  Craswell}{2017}]%
        {10.1145/3020165.3020183}
\bibfield{author}{\bibinfo{person}{Filip Radlinski} {and} \bibinfo{person}{Nick
  Craswell}.} \bibinfo{year}{2017}\natexlab{}.
\newblock \showarticletitle{A Theoretical Framework for Conversational Search}.
  In \bibinfo{booktitle}{\emph{Proceedings of the 2017 Conference on Conference
  Human Information Interaction and Retrieval, {CHIIR} 2017}}.
  \bibinfo{pages}{117--126}.
\newblock


\bibitem[\protect\citeauthoryear{Reddy, Chen, and Manning}{Reddy
  et~al\mbox{.}}{2019}]%
        {reddy-etal-2019-coqa}
\bibfield{author}{\bibinfo{person}{Siva Reddy}, \bibinfo{person}{Danqi Chen},
  {and} \bibinfo{person}{Christopher~D. Manning}.}
  \bibinfo{year}{2019}\natexlab{}.
\newblock \showarticletitle{CoQA: {A} Conversational Question Answering
  Challenge}.
\newblock \bibinfo{journal}{\emph{Trans. Assoc. Comput. Linguistics}}
  \bibinfo{volume}{7} (\bibinfo{year}{2019}), \bibinfo{pages}{249--266}.
\newblock


\bibitem[\protect\citeauthoryear{Ren, Chen, Monz, Ma, and de~Rijke}{Ren
  et~al\mbox{.}}{2020a}]%
        {ren-2020-thinking}
\bibfield{author}{\bibinfo{person}{Pengjie Ren}, \bibinfo{person}{Zhumin Chen},
  \bibinfo{person}{Christof Monz}, \bibinfo{person}{Jun Ma}, {and}
  \bibinfo{person}{Maarten de Rijke}.} \bibinfo{year}{2020}\natexlab{a}.
\newblock \showarticletitle{Thinking Globally, Acting Locally: Distantly
  Supervised Global-to-Local Knowledge Selection for Background Based
  Conversation}. In \bibinfo{booktitle}{\emph{The Thirty-Fourth {AAAI}
  Conference on Artificial Intelligence, {AAAI} 2020}}.
  \bibinfo{pages}{8697--8704}.
\newblock


\bibitem[\protect\citeauthoryear{Ren, Chen, Ren, Kanoulas, Monz, and
  de~Rijke}{Ren et~al\mbox{.}}{2020b}]%
        {ren-2020-conversations-arxiv}
\bibfield{author}{\bibinfo{person}{Pengjie Ren}, \bibinfo{person}{Zhumin Chen},
  \bibinfo{person}{Zhaochun Ren}, \bibinfo{person}{Evangelos Kanoulas},
  \bibinfo{person}{Christof Monz}, {and} \bibinfo{person}{Maarten de Rijke}.}
  \bibinfo{year}{2020}\natexlab{b}.
\newblock \showarticletitle{Conversations with Search Engines: SERP-based
  Conversational Response Generation}.
\newblock \bibinfo{journal}{\emph{arXiv preprint arXiv:2004.14162}}
  (\bibinfo{year}{2020}).
\newblock


\bibitem[\protect\citeauthoryear{Saha, Pahuja, Khapra, Sankaranarayanan, and
  Chandar}{Saha et~al\mbox{.}}{2018}]%
        {DBLP:conf/aaai/SahaPKSC18}
\bibfield{author}{\bibinfo{person}{Amrita Saha}, \bibinfo{person}{Vardaan
  Pahuja}, \bibinfo{person}{Mitesh~M. Khapra}, \bibinfo{person}{Karthik
  Sankaranarayanan}, {and} \bibinfo{person}{Sarath Chandar}.}
  \bibinfo{year}{2018}\natexlab{}.
\newblock \showarticletitle{Complex Sequential Question Answering: Towards
  Learning to Converse Over Linked Question Answer Pairs with a Knowledge
  Graph}. In \bibinfo{booktitle}{\emph{Proceedings of the Thirty-Second AAAI
  Conference on Artificial Intelligence, {AAAI} 2018}}.
  \bibinfo{pages}{705--713}.
\newblock


\bibitem[\protect\citeauthoryear{See, Liu, and Manning}{See
  et~al\mbox{.}}{2017}]%
        {see-etal-2017-get}
\bibfield{author}{\bibinfo{person}{Abigail See}, \bibinfo{person}{Peter~J.
  Liu}, {and} \bibinfo{person}{Christopher~D. Manning}.}
  \bibinfo{year}{2017}\natexlab{}.
\newblock \showarticletitle{Get To The Point: Summarization with
  Pointer-Generator Networks}. In \bibinfo{booktitle}{\emph{Proceedings of the
  55th Annual Meeting of the Association for Computational Linguistics, {ACL}
  2017}}. \bibinfo{pages}{1073--1083}.
\newblock


\bibitem[\protect\citeauthoryear{Sun and Zhang}{Sun and Zhang}{2018}]%
        {10.1145/3209978.3210002}
\bibfield{author}{\bibinfo{person}{Yueming Sun} {and} \bibinfo{person}{Yi
  Zhang}.} \bibinfo{year}{2018}\natexlab{}.
\newblock \showarticletitle{Conversational Recommender System}. In
  \bibinfo{booktitle}{\emph{The 41st International ACM SIGIR Conference on
  Research and Development in Information Retrieval, {SIGIR} 2018}}.
  \bibinfo{pages}{235--244}.
\newblock


\bibitem[\protect\citeauthoryear{Thomas, Czerwinski, McDuff, Craswell, and
  Mark}{Thomas et~al\mbox{.}}{2018}]%
        {10.1145/3176349.3176388}
\bibfield{author}{\bibinfo{person}{Paul Thomas}, \bibinfo{person}{Mary
  Czerwinski}, \bibinfo{person}{Daniel~J. McDuff}, \bibinfo{person}{Nick
  Craswell}, {and} \bibinfo{person}{Gloria Mark}.}
  \bibinfo{year}{2018}\natexlab{}.
\newblock \showarticletitle{Style and Alignment in Information-Seeking
  Conversation}. In \bibinfo{booktitle}{\emph{Proceedings of the 2018
  Conference on Human Information Interaction and Retrieval, {CHIIR} 2018}}.
  \bibinfo{pages}{42--51}.
\newblock


\bibitem[\protect\citeauthoryear{Trippas, Spina, Cavedon, Joho, and
  Sanderson}{Trippas et~al\mbox{.}}{2018a}]%
        {10.1145/3176349.3176387}
\bibfield{author}{\bibinfo{person}{Johanne~R. Trippas},
  \bibinfo{person}{Damiano Spina}, \bibinfo{person}{Lawrence Cavedon},
  \bibinfo{person}{Hideo Joho}, {and} \bibinfo{person}{Mark Sanderson}.}
  \bibinfo{year}{2018}\natexlab{a}.
\newblock \showarticletitle{Informing the Design of Spoken Conversational
  Search: Perspective Paper}. In \bibinfo{booktitle}{\emph{Proceedings of the
  2018 Conference on Human Information Interaction and Retrieval, {CHIIR}
  2018}}. \bibinfo{pages}{32--41}.
\newblock


\bibitem[\protect\citeauthoryear{Trippas, Spina, Cavedon, Joho, and
  Sanderson}{Trippas et~al\mbox{.}}{2018b}]%
        {Trippas:2018:IDS:3176349.3176387}
\bibfield{author}{\bibinfo{person}{Johanne~R. Trippas},
  \bibinfo{person}{Damiano Spina}, \bibinfo{person}{Lawrence Cavedon},
  \bibinfo{person}{Hideo Joho}, {and} \bibinfo{person}{Mark Sanderson}.}
  \bibinfo{year}{2018}\natexlab{b}.
\newblock \showarticletitle{Informing the Design of Spoken Conversational
  Search: Perspective Paper}. In \bibinfo{booktitle}{\emph{Proceedings of the
  2018 Conference on Human Information Interaction and Retrieval, {CHIIR}
  2018}}. \bibinfo{pages}{32--41}.
\newblock


\bibitem[\protect\citeauthoryear{Trippas, Spina, Thomas, Sanderson, Joho, and
  Cavedon}{Trippas et~al\mbox{.}}{2020}]%
        {TRIPPAS2020102162}
\bibfield{author}{\bibinfo{person}{Johanne~R. Trippas},
  \bibinfo{person}{Damiano Spina}, \bibinfo{person}{Paul Thomas},
  \bibinfo{person}{Mark Sanderson}, \bibinfo{person}{Hideo Joho}, {and}
  \bibinfo{person}{Lawrence Cavedon}.} \bibinfo{year}{2020}\natexlab{}.
\newblock \showarticletitle{Towards a Model for Spoken Conversational Search}.
\newblock \bibinfo{journal}{\emph{Inf. Process. Manag.}}  \bibinfo{volume}{57}
  (\bibinfo{year}{2020}), \bibinfo{pages}{102162}.
\newblock


\bibitem[\protect\citeauthoryear{Vaswani, Shazeer, Parmar, Uszkoreit, Jones,
  Gomez, Kaiser, and Polosukhin}{Vaswani et~al\mbox{.}}{2017}]%
        {10.5555/3295222.3295349}
\bibfield{author}{\bibinfo{person}{Ashish Vaswani}, \bibinfo{person}{Noam
  Shazeer}, \bibinfo{person}{Niki Parmar}, \bibinfo{person}{Jakob Uszkoreit},
  \bibinfo{person}{Llion Jones}, \bibinfo{person}{Aidan~N. Gomez},
  \bibinfo{person}{Lukasz Kaiser}, {and} \bibinfo{person}{Illia Polosukhin}.}
  \bibinfo{year}{2017}\natexlab{}.
\newblock \showarticletitle{Attention is All you Need}. In
  \bibinfo{booktitle}{\emph{Proceedings of the 31st International Conference on
  Neural Information Processing Systems, {NIPS} 2017}}.
  \bibinfo{pages}{5998--6008}.
\newblock


\bibitem[\protect\citeauthoryear{Vinyals, Fortunato, and Jaitly}{Vinyals
  et~al\mbox{.}}{2015}]%
        {10.5555/2969442.2969540}
\bibfield{author}{\bibinfo{person}{Oriol Vinyals}, \bibinfo{person}{Meire
  Fortunato}, {and} \bibinfo{person}{Navdeep Jaitly}.}
  \bibinfo{year}{2015}\natexlab{}.
\newblock \showarticletitle{Pointer Networks}. In
  \bibinfo{booktitle}{\emph{Proceedings of 28th Annual Conference on Neural
  Information Processing Systems, {NIPS}, 2015}}. \bibinfo{pages}{2692--2700}.
\newblock


\bibitem[\protect\citeauthoryear{Voskarides, Li, Ren, Kanoulas, and
  de~Rijke}{Voskarides et~al\mbox{.}}{2020}]%
        {10.1145/3397271.3401130}
\bibfield{author}{\bibinfo{person}{Nikos Voskarides}, \bibinfo{person}{Dan Li},
  \bibinfo{person}{Pengjie Ren}, \bibinfo{person}{Evangelos Kanoulas}, {and}
  \bibinfo{person}{Maarten de Rijke}.} \bibinfo{year}{2020}\natexlab{}.
\newblock \showarticletitle{Query Resolution for Conversational Search with
  Limited Supervision}. In \bibinfo{booktitle}{\emph{Proceedings of the 43rd
  International {ACM} {SIGIR} conference on research and development in
  Information Retrieval, {SIGIR} 2020}}. \bibinfo{pages}{921--930}.
\newblock


\bibitem[\protect\citeauthoryear{Vtyurina, Savenkov, Agichtein, and
  Clarke}{Vtyurina et~al\mbox{.}}{2017}]%
        {Vtyurina:2017:ECS:3027063.3053175}
\bibfield{author}{\bibinfo{person}{Alexandra Vtyurina}, \bibinfo{person}{Denis
  Savenkov}, \bibinfo{person}{Eugene Agichtein}, {and} \bibinfo{person}{Charles
  L.~A. Clarke}.} \bibinfo{year}{2017}\natexlab{}.
\newblock \showarticletitle{Exploring Conversational Search With Humans,
  Assistants, and Wizards}. In \bibinfo{booktitle}{\emph{Proceedings of the
  2017 CHI Conference on Human Factors in Computing Systems, {CHI}, 2017}}.
  \bibinfo{pages}{2187--2193}.
\newblock


\bibitem[\protect\citeauthoryear{White and Roth}{White and Roth}{2009}]%
        {DBLP:series/synthesis/2009White}
\bibfield{author}{\bibinfo{person}{Ryen~W. White} {and}
  \bibinfo{person}{Resa~A. Roth}.} \bibinfo{year}{2009}\natexlab{}.
\newblock \bibinfo{booktitle}{\emph{Exploratory Search: Beyond the
  Query-Response Paradigm}}.
\newblock


\bibitem[\protect\citeauthoryear{Wu, Guo, Zhou, Wu, Zhang, Lian, and Wang}{Wu
  et~al\mbox{.}}{2019a}]%
        {wu-etal-2019-proactive}
\bibfield{author}{\bibinfo{person}{Wenquan Wu}, \bibinfo{person}{Zhen Guo},
  \bibinfo{person}{Xiangyang Zhou}, \bibinfo{person}{Hua Wu},
  \bibinfo{person}{Xiyuan Zhang}, \bibinfo{person}{Rongzhong Lian}, {and}
  \bibinfo{person}{Haifeng Wang}.} \bibinfo{year}{2019}\natexlab{a}.
\newblock \showarticletitle{Proactive Human-Machine Conversation with Explicit
  Conversation Goal}. In \bibinfo{booktitle}{\emph{Proceedings of the 57th
  Conference of the Association for Computational Linguistics, {ACL} 2019}}.
  \bibinfo{pages}{3794--3804}.
\newblock


\bibitem[\protect\citeauthoryear{Wu and Yan}{Wu and Yan}{2019}]%
        {wu-yan-2018-deep}
\bibfield{author}{\bibinfo{person}{Wei Wu} {and} \bibinfo{person}{Rui Yan}.}
  \bibinfo{year}{2019}\natexlab{}.
\newblock \showarticletitle{Deep Chit-Chat: Deep Learning for Chatbots}. In
  \bibinfo{booktitle}{\emph{Proceedings of the 42nd International {ACM} {SIGIR}
  Conference on Research and Development in Information Retrieval, {SIGIR}
  2019}}. \bibinfo{pages}{1413--1414}.
\newblock


\bibitem[\protect\citeauthoryear{Wu, Mao, Liu, Zhang, and Ma}{Wu
  et~al\mbox{.}}{2019b}]%
        {10.1145/3331184.3331233}
\bibfield{author}{\bibinfo{person}{Zhijing Wu}, \bibinfo{person}{Jiaxin Mao},
  \bibinfo{person}{Yiqun Liu}, \bibinfo{person}{Min Zhang}, {and}
  \bibinfo{person}{Shaoping Ma}.} \bibinfo{year}{2019}\natexlab{b}.
\newblock \showarticletitle{Investigating Passage-level Relevance and Its Role
  in Document-level Relevance Judgment}. In
  \bibinfo{booktitle}{\emph{Proceedings of the 42nd International {ACM} {SIGIR}
  Conference on Research and Development in Information Retrieval, {SIGIR}
  2019}}. \bibinfo{pages}{605--614}.
\newblock


\bibitem[\protect\citeauthoryear{Zamani and Craswell}{Zamani and
  Craswell}{2020}]%
        {10.1145/3397271.3401415}
\bibfield{author}{\bibinfo{person}{Hamed Zamani} {and} \bibinfo{person}{Nick
  Craswell}.} \bibinfo{year}{2020}\natexlab{}.
\newblock \showarticletitle{Macaw: An Extensible Conversational Information
  Seeking Platform}. In \bibinfo{booktitle}{\emph{Proceedings of the 43rd
  International {ACM} {SIGIR} conference on research and development in
  Information Retrieval, {SIGIR} 2020}}. \bibinfo{pages}{2193--2196}.
\newblock


\bibitem[\protect\citeauthoryear{Zamani, Dumais, Craswell, Bennett, and
  Lueck}{Zamani et~al\mbox{.}}{2020}]%
        {Hamedwww2020}
\bibfield{author}{\bibinfo{person}{Hamed Zamani}, \bibinfo{person}{Susan~T.
  Dumais}, \bibinfo{person}{Nick Craswell}, \bibinfo{person}{Paul~N. Bennett},
  {and} \bibinfo{person}{Gord Lueck}.} \bibinfo{year}{2020}\natexlab{}.
\newblock \showarticletitle{Generating Clarifying Questions for Information
  Retrieval}. In \bibinfo{booktitle}{\emph{Proceedings of the Web Conference,
  {WWW} 2020}}. \bibinfo{pages}{418--428}.
\newblock


\bibitem[\protect\citeauthoryear{Zhou, Zheng, Huang, Huang, and Zhu}{Zhou
  et~al\mbox{.}}{2020}]%
        {zhou-etal-2020-kdconv}
\bibfield{author}{\bibinfo{person}{Hao Zhou}, \bibinfo{person}{Chujie Zheng},
  \bibinfo{person}{Kaili Huang}, \bibinfo{person}{Minlie Huang}, {and}
  \bibinfo{person}{Xiaoyan Zhu}.} \bibinfo{year}{2020}\natexlab{}.
\newblock \showarticletitle{KdConv: {A} Chinese Multi-domain Dialogue Dataset
  Towards Multi-turn Knowledge-driven Conversation}. In
  \bibinfo{booktitle}{\emph{Proceedings of the 58th Annual Meeting of the
  Association for Computational Linguistics, {ACL} 2020}}.
  \bibinfo{pages}{7098--7108}.
\newblock


\end{thebibliography}

\end{document}